\documentclass{emulateapj}
\usepackage{amsmath, amsfonts, amssymb, fancyhdr}

\usepackage{natbib}

\newcommand{\teff}{$T_{\rm eff}$}

\begin{document}
\title{Experimental Design for the Gemini Planet Imager}
\author{James McBride\altaffilmark{1}, James R. Graham\altaffilmark{1, 2}, Bruce Macintosh\altaffilmark{3}, Steven V. W. Beckwith\altaffilmark{1}, Christian Marois\altaffilmark{4}, Lisa A. Poyneer\altaffilmark{3}, Sloane J. Wiktorowicz\altaffilmark{1}} \thanks{E-mail: jmcbride@astro.berkeley.edu}
\altaffiltext{1}{Astronomy Department, 601 Campbell Hall, The University of California, Berkeley, CA}
\altaffiltext{2}{Dunlap Institute for Astronomy \& Astrophysics, University of Toronto, Toronto, Ontario}
\altaffiltext{3}{Lawrence Livermore National Laboratory, 7000 East Avenue, Livermore, CA}
\altaffiltext{4}{National Research Council Canada, Herzberg Institute of Astrophysics, Victoria}
\begin{abstract}
The Gemini Planet Imager (GPI) is a high performance adaptive optics system
being designed and built for the Gemini Observatory. GPI 
is optimized for high contrast imaging, combining
precise and accurate wavefront control, diffraction suppression, and a
speckle-suppressing science camera with integral field and polarimetry
capabilities. The primary science goal for GPI is the direct detection and
characterization of young, Jovian-mass exoplanets. For plausible 
assumptions about the distribution of gas giant properties at large
semi-major axes, GPI will be capable of detecting more than 10\% of
gas giants more massive than 0.5 $M_J$ around stars younger than
100 Myr and nearer than 75 parsecs. For systems younger
than 1 Gyr, gas giants more massive than 8 $M_J$ and with semi-major axes
greater than 15 AU are detected with completeness greater than 50\%. A
survey targeting young stars in the solar neighborhood will
help determine the formation mechanism of gas giant planets by studying
them at ages where planet brightness depends upon formation mechanism. 
Such a survey will also be sensitive to planets at semi-major
axes comparable to the gas giants in our own solar system. In the
simple, and idealized, situation in which planets formed by either the
``hot-start'' model of \citet{Burrows2003} or the core accretion model of
\citet{Marley2007}, a few tens of detected planets are sufficient to 
distinguish how planets form.
\end{abstract}

\section{Introduction} \label{intro_sect}
The past decade has seen extraordinary progress in detection of
exoplanets with the deployment of custom designed systems that use
the Doppler method, such as HARPS \citep{Mayor2003}, 
and the transit method, such as Kepler \citep{Borucki2003}.
Exoplanetary systems are also being discovered using microlensing 
\citep{Gould2010, Sumi2010}.
No planets are confirmed to have been detected using astrometry, though there 
are disputed claims of some detections, including a companion to VB 10 
\citep{Pravdo2009, Bean2010, AngladaEscude2010}.
Even so, the upcoming Gaia mission is expected to have astrometric precision
of $\sim 8 \mu$as, and thus be able to astrometrically discover the
hundreds of exoplanets and exoplanetary systems expected to have astrometric 
signals $> 30\; \mu$as around stars within 200 parsecs \citep{Casertano2008}. 
Polarimetric detection of HD 189733 b has also been claimed by 
\citet{Berdyugina2008}, but \citet{Wiktorowicz2009} was unable to reproduce 
the detection. 

These methods are indirect, and provide different information about
the planets that they detect. 
From radial velocity surveys alone, the period, eccentricity and a minimum 
mass may be measured for detected planets. 
For planets that also transit their hosts, 
it is possible to learn the inclination of the orbit and planet radius,
given simple assumptions about the system and knowledge of the stellar
mass-radius relation, which then give the mass and density of the 
planet \citep{Seager2003}.
Mid-IR exoplanetary light has been detected in secondary eclipses by
Spitzer \citep{Charbonneau2005, Grillmair2007, Richardson2007, 
Fressin2010, O'Donovan2010}, though information about the planet spectrum 
itself is limited by the photon shot noise of the primary.

\citet{Marcy2008} lists 228 extrasolar planets, with 5\% of targeted stars
possessing massive planets, and shows that a diversity of
exoplanet systems exists. Though the number of radial velocity confirmed
planets has nearly doubled 
since,\footnote{The Exoplanet Orbit Database at http://exoplanets.org 
lists 429 planets as of March 2010 \citep{Wright2010}.} and many more
planet candidates exist in the first four months of Kepler data 
\citep{Borucki2011}, 
radial velocity and transit surveys leave several long-standing
questions about planetary systems unanswered. How do planets form? Is
the solar system typical? What is the abundance of solar-like systems?
These surveys also raise new questions, including what
produces the dynamical diversity in exoplanetary systems?

The prospects for indirect planet detection techniques alone to answer these
questions is limited. With the exception of microlensing,
the effectiveness of these techniques at
detecting exoplanets decreases at large semi-major axes. For a reliable 
detection with radial velocity, 
a significant fraction of an orbital period must elapse. 
As such, radial velocity surveys are only now 
reaching the precision and lifetime
necessary to detect Jupiter, and thus do not yet constrain the
frequency of solar system analogs. For example, in
the Keck radial velocity search that began in 1996 July \citep{Butler2006},
only planets with $a \leq 6$ AU have completed one
orbit. The median semi-major axis of known exoplanet orbits is approximately
1 AU, and HD 190984 b is the only planet with 
with $a > 6$ AU \citep{Santos2010}.
As the time baseline of surveys grows, the semi-major axes probed
increase according to $a = P^{2/3}$, meaning improving the statistics 
at large semi-major axes will be challenging. Doppler surveys have also
been focused on a small range of stellar masses and ages, as precise radial
velocity measurement requires strong stellar spectral features 
and low activity. This has limited
studies of exoplanet trends with stellar properties beyond
basic correlations between host star mass and metallicity and the 
probability of hosting a planet within a few AU \citep{Johnson2010}.

Direct imaging constitutes the next step in characterizing other planetary
systems. Many previous attempts to directly image substellar companions 
to stars have already placed stringent constraints on the presence of
giant planets on wide orbits \citep{Schroeder2000, Masciadri2005, 
Kasper2007, Biller2007, Lafreniere2007b, Apai2008, Heinze2010, Leconte2010}. 
There are also projects already underway, such as 
NICI \citep{Liu2010} and SEEDS \citep{Tamura2009}, 
or that are near deploying, such as SPHERE \citep{Beuzit2010} and GPI. 
Thus far, planet detections through direct imaging have been 
limited by difficulty in achieving high
contrast at small angular separations, but instruments are now reaching
the threshold at which planet detection by direct imaging is promising.
Direct imaging is highly complementary to Doppler surveys.
Unlike Doppler searches, direct imaging is sensitive at large
semi-major axes, as planets 
can be found without waiting for an orbit to
complete---a condition that renders Doppler detection of planets on 
Neptune-like orbits with semi-major axes of 30 AU and periods of 160 yr
impractical. Fourier decomposition, which underlies Doppler and
astrometric detection, is also subject to aliasing and beat
phenomena, and suffers confusion when multiple planets are
present. A direct search that can probe beyond 5 AU will thus bring 
statistical significance to these studies. 

The goal of direct detection is to spatially separate the exoplanet light
from that of its primary. This affords access to exoplanet atmospheres, which
yields fundamental information including effective temperature,
gravity, atmospheric composition and abundances, orbital motion, and
perhaps even weather and planetary spin 
(via polarization associated with rotation-induced
oblateness; \citealt{Marley2010}). Direct imaging may also reveal 
the role of giant impacts, though 10 to
100 planets $\sim$ 10 Myr old may need to be discovered in order to
expect to find a single gas giant post impact \citep{Anic2007},
or lead to discovery of time dependent atmospheric phenomena. 

Most importantly, by probing large semi-major axes, direct imaging will see 
gas giants beyond the ``snow line,'' which is where they are expected to
form, and extend out to the greatest distances at which giant planets can form.
The location of the region of interest depends
on at least two competing factors: time-scales for planet building and
the availability of raw material. Dynamical and viscous time scales in
the disk are shorter at small radii, while for typical surface-density
laws the amount of mass increases with radius, with a jump in the
abundance of solid material beyond the ``snow line'' where ices
condense. This change in the surface density of solid material occurs
at 2.7 AU in the Hayashi model \citep{Hayashi1981}. The location of this
boundary depends on the disk structure \citep{Sasselov2000}, but
for solar type stars, the zone of interest is beyond that which is
readily probed by the Doppler method. The discovery of giant planets
far beyond the snow line would tend to favor theories of planet
formation by gravitational instability over solid core condensation
and accretion. At larger orbital radii ($>$ 20--30 AU), gas-cooling times
become shorter than the Keplerian shearing time 
(e.g., \citealt{Kratter2010})---a necessary condition
for runaway gravitational instability \citep{Gammie2001, Johnson2003,
Boss2002}---while solid core growth by collisional coagulation
of planetesimals proceeds prohibitively slowly \citep{Goldreich2004}.
Even so, new ideas about coupling between migration and core accretion 
suggest significantly increased growth, allowing the formation of planets well
beyond the snow line \citep{Levison2010}.
The structure of our own solar system implies that a full
picture of planet formation cannot be constructed without reaching out
to 30 or 40 AU. Millimeter observations of T Tauri disks support this, 
as typical
disk radii fall in the range 50--100 AU \citep{Isella2009}. 

The initial conditions, composition, and equation of state all influence 
evolution of young gas giant planets. Gas giants will take tens 
to hundreds of millions of years to ``forget'' 
their post-formation entropy, meaning 
temperatures and luminosities of young planets will reveal their past
\citep{Marley2007}. Stellar properties, in particular mass and metallicity,
will also influence planet formation. The observed planet-metallicity
correlation in Doppler planets has been interpreted as supporting
core accretion as the formation mechanism of gas giants \citep{Fischer2005, 
Johnson2010}.
Metal-rich disks are expected to have longer lifetimes \citep{Ercolano2010} 
and enhanced clump formation \citep{Johansen2009}, all increasing the 
likelihood of planet formation for those formed 
by core accretion. Formation by disk instability
is not expected to be sensitive to stellar metallicity \citep{Boss2002}. 
This has led to the suggestion that low metallicity stars will 
preferentially form planets by disk instability on wide orbits, while
high metallicity stars will host planets on short period orbits formed by
core accretion \citep{Boley2009, DodsonRobinson2009}.
Stellar metallicity is also expected to influence planet appearance. 
Planets formed via core accretion are predicted to themselves have
super-stellar metallicity \citep{Pollack1996, Fortney2008}. Though 
gas giants more massive than $\sim 5 M_J$ formed by disk instability
may also have super-stellar atmospheric metallicities,
less massive gas giants should have metal depleted atmospheres 
\citep{Helled2008}. 
Exoplanet infrared colors are predicted to be sensitive to 
atmospheric metal abundances, so direct imaging and spectroscopy
will reveal information about a planet's formation \citep{Fortney2008}.

Another reason to image the outer regions of extrasolar systems is to
probe them for vestiges of planetary migration. Ninety percent of the
Doppler sample consists of massive planets with $a <$ 3 AU, suggesting
that they migrated inwards to their present locations. A variety of
mechanisms may drive orbital evolution; the tidal gravitational
interaction between the planet and a viscous disk \citep{Goldreich1980},
the gravitational interaction between two or more
Jupiter mass planets \citep{Rasio1996}, and the interaction between
a planet and a planetesimal disk \citep{Murray1998}. It is
energetically favorable for a Keplerian disk to evolve by transporting
mass inward and angular momentum outward \citep{Lynden-Bell1974}.
Consequently, inward planetary drift appears inevitable, and
this is what is found in certain simulations \citep{Trilling2002,
Armitage2002, Matsuyama2003}. However,
if planets form while the disk is being dispersed, or if multiple
planets are present, outward migration can also occur. In a system
consisting initially of two Jupiter-like planets, a dynamical
instability may eject one planet while the other is left in a tight,
eccentric orbit. The second planet is not always lost; the observed
Doppler exoplanet eccentricity distribution can be reproduced if the
51 Pegasi systems are formed by planet-planet
scattering events and the second planet typically remains bound in a
wide ($a >$ 20 AU), eccentric orbit \citep{Rasio1996, Marzari2002, Veras2009}.
Divergent migration of pairs of Jupiter-mass
planets within viscous disks leads to mutual resonance crossings and
excitation of orbital eccentricities such that the resultant
ellipticities are inversely correlated with planet masses \citep{Chiang2002}.
Given decreasing disk viscosity with radius
and the consequent reduction in planetary mobility with radius, we
expect eccentricities to decrease with radius, perhaps sharply if the
magneto-rotational instability is invoked \citep{Sano2000}. By
contrast, excitation of eccentricity by disk-planet interactions
requires no additional planet to explain the ellipticities of
currently known solitary planets \citep{Goldreich2003}. Clearly,
observations of the incidence, mass, and eccentricity distributions of
multiple planet systems would sharpen ideas regarding how
planetary orbits are sculpted.

There have already been a number of planets and 
planet candidates discovered by direct imaging, 
including beta Pic b (8 AU/8 $M_J$; \citealt{Lagrange2009}),
the upper Sco object 1RXS J160929.1-210524 
(330 AU/8 $M_J$ \citealt{Lafreniere2008}), HR 8799 b, c, d, and e
 (24, 38, 68, and 14 AU/10, 10, 7, and 10 $M_J$; 
\citealt{Marois2008a, Marois2010b}), 
and Fomalhaut b (120 AU/$<$ 2 $M_J$ \citealt{Kalas2008}).  
While the sample of exoplanets is incomplete for $a >$ 5 AU,
indirect searches continue to hint that the semi-major axis
distribution is at least flat, and possibly rising, in $dN/d\log(a)$
beyond 5 AU \citep{Cumming2008}. 
The sample of six microlensed planets beyond the
ice line supports this trend \citep{Sumi2010, Gould2010}.
Thus, a direct imaging search of outer solar system regions (5--50 AU), such as
proposed here, would increase the total number of planets found
relative to those in inner solar system orbits ($<5$ AU). 
The goal of direct imaging is to assemble the first statistically
significant sample of exoplanets that probes beyond the reach of
indirect searches and quantifies the abundance of solar systems like
our own.

\section{Overview of the GPI instrument} \label{inst_sect}
The Gemini Planet Imager is configured to allow high contrast ($> 10^7$) 
imaging on angular scales 
of the diffraction limit ($\sim 5-20 \lambda / D$). High-contrast imaging 
with current AO systems is almost completely limited by quasi-static 
artifacts caused by slowly evolving wavefront errors. These originate 
from many sources, including inadequately calibrated non-common-path 
errors that arise from the difference between the science and wavefront 
sensing paths, uncorrectable high spatial frequency errors on instrumental 
and telescope optics, aliased high-frequency wavefront 
errors \citep{Poyneer2004}, chromatic errors, reflectivity variations, 
and Fresnel propagation effects. Although these can be partially 
attenuated through techniques like angular differential 
imaging \citep{Marois2006}, the errors evolve rapidly enough 
that ADI-like techniques do not operate well at small angles 
where it takes significant time for enough field rotation to 
accumulate. Unfortunately, these small angles are also the scales 
that correspond to our own solar system for most nearby young stars. 
GPI was designed from the beginning to minimize these quasi-static 
error sources and hence allow high-contrast imaging approaching the 
photon noise limit. For more detail, see \citet{Macintosh2006, Macintosh2008}.

The GPI system consists of five key subsystems: an AO system, a coronagraph, an interferometer, an integral field spectrograph, and a software system.

The AO system makes fast measurements of the
instantaneous wavefront and provides wavefront control via 
a tip-tilt stage and two deformable mirrors, one conventional 
piezoelectric and one high-order silicon micro-electro-mechanical-system 
(MEMS) device. The high-order AO system has 43 actuators across the 
diameter of the 7.8-m Gemini South primary mirror and operates at 
at an update rate of 1.5 kHz. The AO system uses a spatially-filtered 
Shack-Hartmann wavefront sensor to minimize aliasing and have uniform 
response in varying atmospheric seeing \citep{Poyneer2004}. The 
anti-aliasing produces a characteristic square ``dark hole'' 
region, $\sim 43 \lambda/D$ on a side. 

The coronagraph is an apodizer-pupil Lyot coronagraph (APLC), and it
controls diffraction \citep{Soummer2006}. This 
combines a carefully-designed input apodization with a focal-plane 
occulting stop (in GPI, a mirror with a central hole) and a Lyot pupil 
stop matched to the input telescope pupil size. The GPI APLC 
implementation is optimized for achromatic performance to improve 
multi-wavelength speckle suppression. 
Each waveband has a hard-edged occulter hole 
with a radius of 2.8 $\lambda/D$, 
but that hole is also surrounded 
by a bright Airy ring, so high contrast is only fully practical 
at ($\sim 4$--$5\lambda/D$).

The interferometer provides low-temporal bandwidth precise and
accurate measurements at the science wavelength of the time-averaged wavefront 
delivered to the coronagraph occulting spot.
This information is used to remove slowly-evolving quasi-static errors 
due to flexure and changes in the response of the main visible-light 
wavefront sensor. 

The science instrument is an integral field spectrograph (IFS)
  that images in simultaneous multiple wavelength channels. 
The lenslet-based IFS has sampling of 0.014$\arcsec$ per lenslet, 
with a square field of view of 2.8$\arcsec$ on a side. Each spatial 
pixel is dispersed into a spectrum with resolution $R\sim 45$. 
A single observation covers one of the $Y$, $J$, or $H$ bands, or  
half of the $K$ band, which is split in to $K_1$ and $K_2$. 
Raw detector images are reassembled by a data pipeline into 
$200 \times 200 \times 16$ element data cubes. The IFS also 
includes a differential polarimetry mode, intended for characterization 
of circumstellar dust and not discussed in this paper.  
Finally, the software system coordinates communication between the other
subsystems and the observatory software.

GPI operates at a Cassegrain focus on the alt/az mount of the Gemini 
telescope in a fixed orientation with respect to the telescope, 
to increase instrumental stability and enable ADI post-processing. 
A typical science observation will consist of an hour long sequence 
of 30-60 second exposures. This is long enough that read noise and 
dark current are unimportant contributors to the noise budget, and 
short enough that image motion during the exposure is minimal. 

\section{Instrumental Noise Calculations} \label{noise_sect}
\subsection{Simulations}
High-contrast imaging noise can be broken down into two main categories: speckle noise and photon noise. Instantaneous monochromatic high-contrast images consist of a pattern of bright ``speckles'' surrounding the central core. These speckles have a size of $\sim \lambda/D$, comparable to the image of a planet, and their random fluctuations are usually the main limitation in planet detection in existing high-contrast imaging instruments. In many cases (e.g. atmospheric turbulence wavefront errors), the speckle pattern will evolve rapidly, and after some characteristic speckle lifetime has elapsed the speckle noise will scale with exposure time as $t^{0.5}$ and in a long exposure produce a more uniform halo of light. In addition, any light present in the focal plane due to wavefront errors, whether speckled or uniform, will contribute Poisson photon noise. 

In the GPI architecture, we consider the following sources of wavefront error and their corresponding speckle and/or photon noise \citep{Poyneer2006a, Marois2008b}:

\begin{enumerate}

\item Residual atmospheric errors. Even operating at 1.5 kHz, 
the motion of the turbulent atmosphere will cause changes in 
wavefront between measurement and the application of a correction. 
This is the dominant source of mid-to-low frequency wavefront errors. 
Here, low-frequency errors ($<3$ cycles/pupil) refers to errors at 
spatial frequencies mostly blocked by the coronagraph occulter, and 
mid-frequency errors (3--22 cycles/pupil)
are those transmitted by the occulter but that are
within the controllable range of the deformable mirror. Errors
at these frequencies contribute scattered light and photon noise near the star.
Higher frequency spatial wavefront errors ($>22$ cycles/pupil) are unsensable and 
uncorrectable by the GPI AO system and contribute light outside the 
dark hole. A representative speckle produced by these 
atmospheric wavefront errors 
has a lifetime of a few tenths of a second and an amplitude of 60 nm. 

\item Wavefront sensor measurement noise. On dimmer stars, the 
individual wavefront sensor measurements will include a significant 
noise component that will translate to random errors of position 
of the deformable mirror. In a Shack-Hartmann wavefront sensor, these 
errors are not spatially white, but have increasing power at low spatial 
frequencies. On dim stars, this is again a significant source of 
scattered light over the whole dark hole region. Speckles associated 
with this error source randomize at the closed-loop bandwidth of the 
system, with lifetimes of a few milliseconds, rapidly smoothing out.
A typical value is 25 nm for a star with $I = 6$ mag.

\item Residual non-common-path wavefront errors. After daytime calibration 
and closed-loop correction by the precision interferometric wavefront 
sensor, these are assumed to be $\sim5$ nm of low-frequency error and 
$\sim1$ nm of mid-frequency errors. The amount of light 
scattered by these small wavefront errors produces a negligible amount 
of photon noise, but the slowly-varying speckles they produce are a
significant source of speckle noise. 

\item Reflectivity variations and amplitude errors. The GPI deformable 
mirrors are operated in phase conjugation mode, correcting the phase 
errors of the wavefront and producing a symmetric final PSF. As a result, 
any light scattered by changes in amplitude---for example, from 
reflectivity variations on the Gemini primary mirror---is uncorrectable. 
We assign a reflectivity variation to each individual optic in our system 
(typically 0.1\%, rising to 1\% for the primary mirror and the MEMS DM, 
with a $k^{-2.5}$ spatial frequency power spectrum). The light scattered 
by these reflectivity errors again produces little photon noise but 
significant speckle noise. 

\item Fresnel propagation errors. Surface errors on an optic at an arbitrary 
location in the GPI optical train will initially produce a pure phase error 
in the wavefront, but as the wavefront propagates these errors will mix 
between phase and amplitude \citep{Shaklan2006}. The finite size of GPI optics 
will also produce amplitude fluctuations near the edge of the beam. As noted 
above, GPI will only correct the component of these errors that is realized 
as phase at the deformable mirrors, not the amplitude component. To mitigate 
this, GPI optics are located away from focal planes and are of very high 
quality ($\sim 2$ nm RMS wavefront error typically), but these are still a 
significant source of persistent and chromatic speckles. 

\item Atmospheric scintillation. Classical scintillation causes uncorrectable 
amplitude fluctuations in the telescope pupil similar to the Talbot propagation 
errors. Unlike the internal Talbot errors, those from the atmosphere are time 
varying. Simulations and calculations show that for our typical atmosphere 
profile this is a negligible contribution to the photon and speckle noise and 
hence it is ignored. 
\end{enumerate}

In addition, normal astronomical noise sources such as sky background, detector 
readout, and dark current will be present, though for typical GPI targets 
with $H < 8$ mag. these are negligible. 
Telescope vibrations (windshake or mechanical vibrations of either GPI or 
the telescope secondary) are not included in this model, in part because 
the exact vibration environment of the Gemini telescope is unknown. The GPI 
tilt error budget allocates $<5$ mas RMS to these effects; GPI employs an 
advanced tip/tilt control algorithm \citep{Poyneer2010} that can cancel 
out resonant vibrations of the telescope, and the mechanical structure 
includes tuned-mass damping to cancel internal vibrations. These 5 mas 
have no significant effect on the coronagraph or on planet detectability.

Modeling all these effects is challenging. Even with simple Fraunhoffer 
propagation, simulating the AO wavefront correction and propagation 
to the science focal plane requires several CPU-hours per second of 
exposure time. Since GPI has been designed to minimize the quasi-static 
wavefront errors 3, 4, and 5 in the list above, 
their effects can only be seen after 
many minutes of integration. As a result, we treat the dynamic and 
static wavefront errors as independent and treat them in two different 
simulations. The first simulates the dynamic behavior of the atmosphere 
and each component of the adaptive optics system to produce 
short-exposure point spread functions, using Fraunhoffer propagation 
through the GPI optics and coronagraph. These were run for a standard 
Gemini Cerro Pachon atmosphere \citep{Tokovinin2006} 
and for stars with brightness 
ranging from $I$ = 5--9 mag. \citep{Poyneer2006a}. 
The second, described in \citet{Marois2008b}, includes all the 
quasi-static wavefront error source and full Fresnel propagation 
through the GPI optical train, but has no atmospheric wavefront errors. 
Both simulations are run at each IFS wavelength channel within the $H$ band. 
Photon and speckle noise are evaluated separately in each simulation. 

\subsection{Post-processing}
Speckle noise can be attenuated through post-processing techniques such as ADI. However, this depends critically on the stability of the aberrations producing those speckles. We have assumed no ADI attenuation whatsoever, making the pessimistic assumption that wavefront errors will evolve just fast enough to be unsubtractable over the timescales needed for ADI field rotation. Field rotation will still produce some averaging of the residual speckle noise as the planetary companions move through the speckle noise pattern, so we reduce the speckle noise by an amount equal to the square root of the number of $\lambda / D$ that field rotation will move the planet through in a 1-hour exposure. 

Since GPI's science instrument produces spectral data cubes, speckles can 
also be attenuated by scaling and subtracting different wavelength channels. 
We assume post-processing using a simple ``double difference'' algorithm 
using three different wavelengths. For the atmospheric speckles, which 
are almost purely phase errors and hence have well-behaved chromaticity, 
this attenuates speckle noise by more than 4 to 6 magnitudes at all radii, 
leaving the atmospheric speckle noise completely negligible. 
For the quasi-static speckles, especially the Fresnel effects, 
the speckles are more chromatic and they are attenuated by only 2--3 
magnitudes within the dark hole region. Since these speckles are 
weak to begin with, the residual noise is lower than the photon 
noise for all but the brightest stars. 

This double-difference subtraction will also attenuate the planetary 
signal, unless the planetary signal contains deep molecular 
absorption features (e.g. methane), or the wavelength scaling 
of the speckles cause them to move by more than $2 \lambda/D$ 
through a single spectral band, i.e. at radii greater 
than $10 \lambda/D$ for a 20\% bandpass. Although not all planets 
will show methane absorption, the faintest (and hence most 
difficult to detect) planets are expected to. The HR 8799 
planetary companions do not show methane absorption but are so 
bright---contrast $\sim 10^5$ relative to their F-star host---that 
they would be easily detected by GPI even without spectral 
differencing. Hence the use of the spectral-difference contrast 
curves is appropriate for the Monte Carlo modeling described here.
In actual operation, GPI will likely use a more sophisticated approach to PSF
subtraction with wavelength and time, likely based on the LOCI algorithm
\citep{Lafreniere2007a} and the SOSIE framework \citep{Marois2010a}.
However, this would not significantly change the results here, since
in these simulations even with simple spectral differencing GPI is
photon-noise rather than speckle-noise limited for stars 
brighter than $I$ = 6 mag.

\subsection{Final instrumental contrast predictions} \label{final_cont_sect}
\begin{figure}[!htb]
\begin{center}
\includegraphics[scale=0.4]{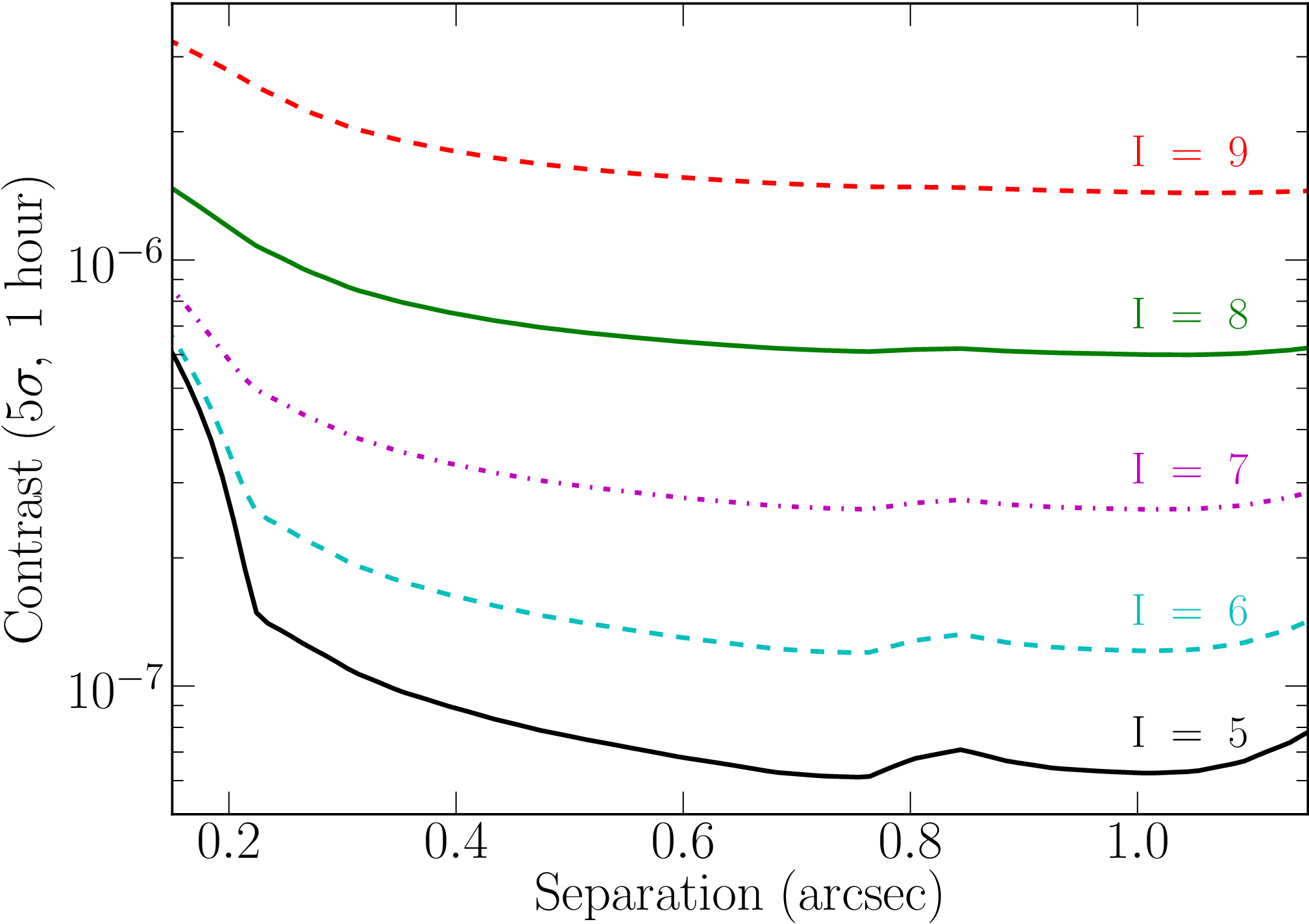}
\caption{Contrast curves (5$\sigma$) in the $H$ band as a function of angular 
separation for five different star $I$ magnitudes, assuming one hour on 
the target and an A0 star. Photon shot noise dominates for stars with 
$I > 6$ mag., while static speckles dominate for stars with $I < 6$ mag. Visit
http://planetimager.org for a tabular version of these data.} 
\label{cont_curves}
\end{center}
\end{figure}

The results of the instrumental noise calculations are tabulated as a
function of stellar brightness and angular separation for each noise source. 
These tabulated values are used to generate the total number of photons
contributed by all sources of noise for some exposure time. A planet
may then be considered detectable if the expected number of photons detected
from the planet is five times the number of photons contributed by noise. 
The contrast
curves in Figure \ref{cont_curves} summarize this result, showing the
ratio of five times the overall noise to the number of photons detected
from the host star as a function of angular separation. The instrument
is expected to reach contrasts better than $10^{-7}$ over an hour long
exposure for targets with $I \sim 5$ mag. For brighter targets, the
performance remains the same, while the expected contrast declines for
dimmer targets. For stars beyond $I \sim$ 10--11 mag., the system performance
becomes comparable to existing instruments. For that reason, stars brighter
than $I \sim 10$ mag. will make up the majority of GPI targets.

\subsection{Matched filters} \label{matched_sect}
\begin{figure}[!htb]
\begin{center}
\includegraphics[scale=0.4]{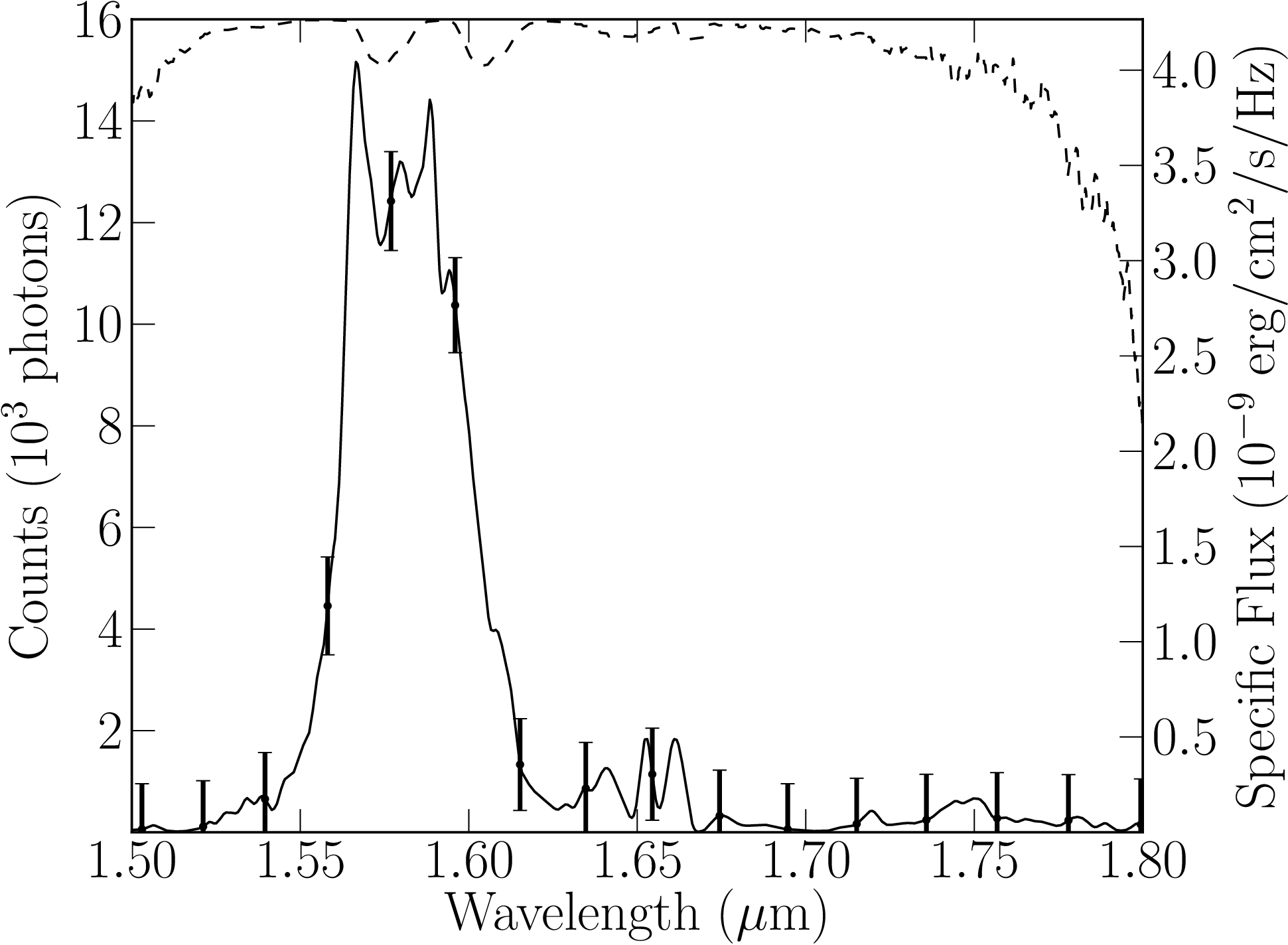}
\caption{A smoothed model spectrum for an atmosphere 
with $T_{\rm eff} = 400$ K and $\log(g) = 3.66$
is shown as a solid black line, with black points indicating
where the spectrum is sampled, and system noise in
each channel as solid black bars. This is representative of a 4 $M_J$ planet
around a 500 Myr Sun-like star 40 parsecs from Earth. The dashed black line
shows terrestrial atmospheric transmission.} \label{noisy_spec}
\end{center}
\end{figure}

Models of the spectral properties of planets (these models will
 be discussed in more detail in the next section)
predict young planets will show many spectral features, 
rather then being uniformly bright across a wavelength band.
For this reason, the integral field spectrograph, which provides 
16 spectral channels across each wavelength band, will be important
to the success of GPI at detecting planets.
A planet may be bright enough in some spectral channels to be above
the contrast limit, while its flux when summed over the entire band is not.
The model spectrum from \citet{Fortney2008} in Figure \ref{noisy_spec} is
an example of such a case. The flux from this model planet
is concentrated in a narrow range of wavelengths, while the
rest of the spectral channels contribute primarily noise. When looking 
at the spectrum, even without the model spectrum to guide the eye,
it looks like a clear detection. Yet the overall signal to noise 
ratio across the entire H band is less than 5, meaning a sum over
all spectral channels would yield a non-detection.

Analysis of real data will involve much more than evaluating a planet's
signal to noise ratio and declaring a detection or non-detection. For the 
purposes of simulating tens of thousands of planets around thousands of stars
though, this is a practical approach.
Yet the spectral features in planetary atmospheres suggest that 
generating a planet's signal to noise ratio by taking a simple sum of 
all spectral channels will underestimate GPI's ability to detect that planet.
To explore how much the detection rate may be affected,
we used model atmospheres to generate matched filters. This was done by
degrading the resolution of a model atmosphere at a given surface gravity and 
temperature to match that of GPI ($\lambda / \Delta \lambda \sim 45$ in 
the $H$ band),
and then weighting each of the spectral channels according to the model
predicted flux. These weighted channels are then  
summed to find the total flux from the planet, and the total noise. 

Using matched filters significantly improves survey performance. 
Relative to a filter that gives equal weight to all spectral channels, 
a matched filter corresponding to an atmosphere at 400 K and surface 
gravity $\log(g) \sim 3.6$ (in cgs) from \citet{Fortney2008}
provides a factor of 3 improvement 
of the detection rate. This assumes that the planet atmospheres are as
described by \citet{Fortney2008} as well, so this represents the ideal
situation. Using the model 
atmospheres of \citet{Burrows2003} to generate a matched filter instead, 
the result changes marginally. 
However, given uncertainties in planet atmospheres, particularly cooler
gas giants, all subsequent simulations results reported here do not 
use matched filters. As a consequence, estimated planet detection rates may be 
somewhat pessimistic.

\section{Planet Models} \label{model_sect}

We consider planets without internal energy sources (i.e., deuterium fusion)
that are young and sufficiently far
from their primary star that we can ignore the stellar radiation field when
considering cooling and contraction, i.e., \teff$ >> (R_*
/a)^{1/2}T_*$, or \teff$^4 >> (180 K)^4$ for a planet orbiting a solar type
star at $a = 5$ AU. 

The luminosity of a planet is given by its radius and effective
temperature, but the detectability at specific wavelengths also
depends on the nature of the dominant opacity sources in the
photosphere. The photospheric chemical composition at a given
effective temperature is sensitive to surface gravity, $g =
GM/R^2$. Both $T_{\rm eff}$ and $\log(g)$ are depend upon the age and the
mass of the planet. To understand the detectability of a planet then
involves converting a planet age and mass to a luminosity,
effective temperature, radius, and surface gravity. 

\begin{figure}[!htb]
\begin{center}
\includegraphics[scale=0.4]{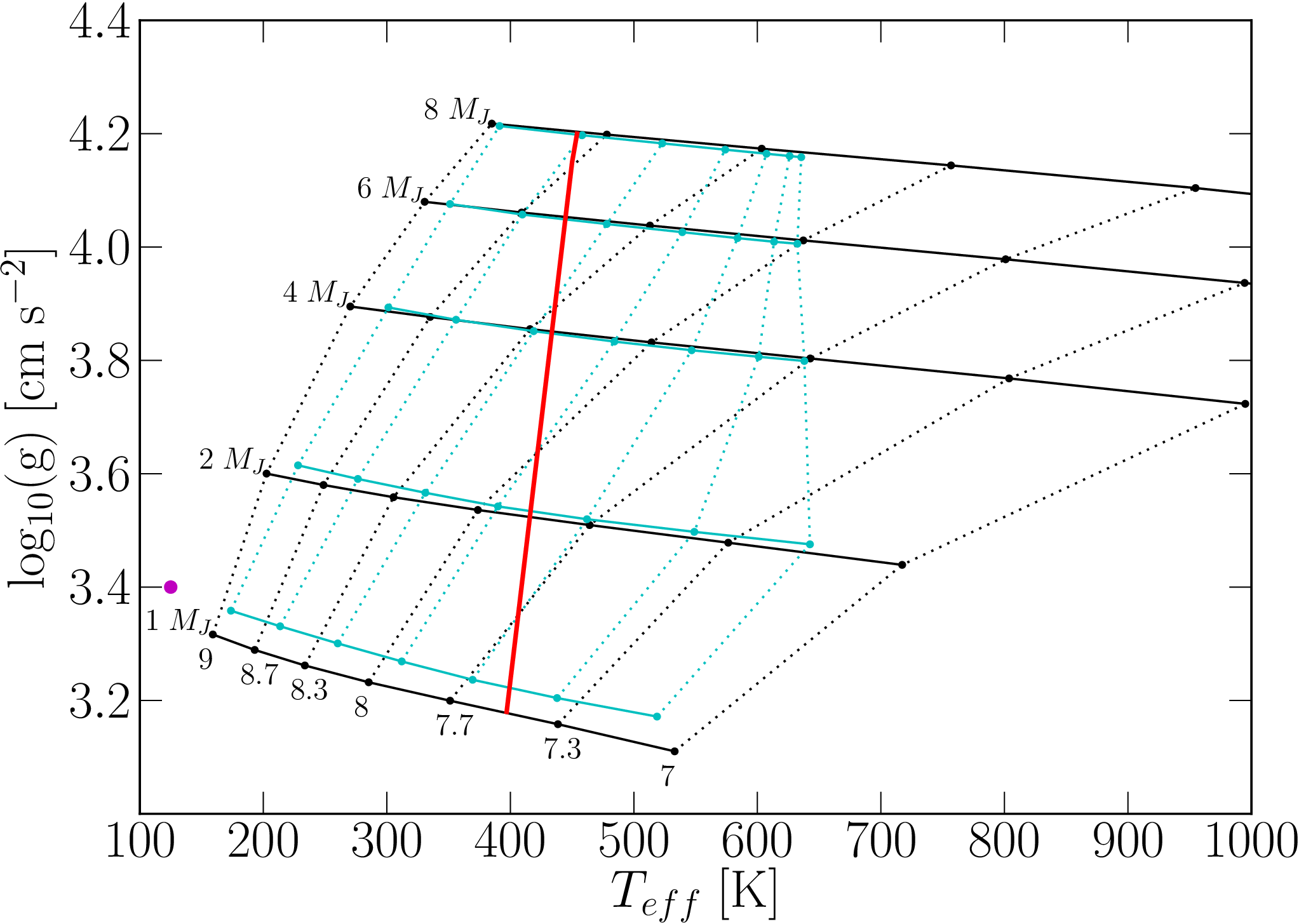}
\caption{Surface gravity plotted against effective temperature for
selected mass and age contours for two planet formation models, where
contours are labeled in $M_J$ and $\log(t)$. The
\citet{Burrows2003} hot start model is shown in black, and the 
\citet{Marley2007} core accretion model is in cyan. 
The H$_2$O condensation line is shown in red, 
below which planet atmosphere calculations become more uncertain. The
position of Jupiter, at $T_{\rm eff} \approx 125$ K, $\log(g) \approx 3.4$, and
$\log(t) \approx$ 9.66, is shown as a magenta dot.
} \label{teff_logg_comp}
\end{center}
\end{figure}

To establish this mapping, we compare two evolutionary models. 
One is the so called ``hot
start'' model of \citet{Burrows1997}, and the other is the the 
core accretion model of \citet{Marley2007}. The most significant
difference between the two is summarized in Figure \ref{teff_logg_comp},
which shows effective temperature and surface gravity for different
mass and age contours.
While the expected effective temperatures and surface gravities of the
two models come in to rough agreement for older planets, planets
that form via core accretion are expected to be significantly cooler
post-formation than planets formed via a hot start. 

\subsection{Hot start models}
The label ``hot start'' was used by \citet{Marley2007} to describe 
models of the atmospheres and evolution of extrasolar giant planets that
ignored the different formation mechanisms of brown dwarfs and gas 
giants. These models assumed that an object's mass, age, and composition
dominated its spectral characteristics, and assumed that gas giant planets,
like brown dwarfs, form like stars. Objects formed
in this manner possess a fully adiabatic interior with high specific 
entropy, which corresponds to a high internal temperature.  
These models focus on treating the atmospheres of gas giant planets in
order to predict the emergent flux as a function of temperature and surface
gravity.

The hot start model used here is that of \citet{Burrows1997},
which spans an effective
temperature range from approximately 100 K to 1000 K, covering the
realm between the known Jovian planets and the known T dwarfs. 
The models can be conveniently approximated as power laws, found by
performing a regression over a restricted range of masses and ages of
interest: 1 $<$ $M/M_J$ $<$ 12 and 0.01 $<$ t/Gyr $<$ 2. The
resultant expressions are 
\begin{eqnarray}
T_{\rm eff}(M, t) & = & (145\;K) \left(\frac{t}{{\rm Gyr}}\right)^{-0.29} \left(\frac{M}{M_J}\right)^{0.47}, \\
L(M, t) & = & (5.4 \times 10^{-9} L_\odot) \left(\frac{t}{{\rm Gyr}}\right)^{-1.21} \left(\frac{M}{M_J}\right)^{1.87}, \\
R(M, t) & = & (1.13 R_J) \left(\frac{t}{{\rm Gyr}}\right)^{-0.034} \left(\frac{M}{M_J}\right)^{-0.013}.
\end{eqnarray}
These expressions are comparable to previous semiempirical estimates of these
dependences \citep{Black1980}, and the luminosity expression is close to that
of cooling at constant heat capacity ($L \sim M^2 t^{-4/3}$).
The resultant radii and effective temperatures have rms errors of 2\% and
4\% respectively. The corresponding error in bolometric magnitudes is
0.15 mag. We use the radiative-convective equilibrium atmosphere model
of \citet{Marley1996}, further described in \citet{Burrows1997};
the most recent tabulations are provided in \citet{Burrows2003}.
The model has been updated to self-consistently include
both alkali opacities \citet{Burrows2000} and precipitating clouds
\citep{Ackerman2001}. 

A major source of uncertainty in planet atmosphere models is the
treatment of clouds. As with the known T and L dwarfs, absorption by
water vapor dominates the spectra of the cooler brown dwarfs. These
features generally deepen with increasing age and decreasing
mass. This trend is in part due to the increase with decreasing
gravity of the column depth of water above the photosphere. At
effective temperatures below 400--500 K, water vapor condenses in
planetary atmospheres. The appearance of water-ice clouds constitutes
a major uncertainly separating the known T dwarfs from the giant
planets, and is denoted in Figure \ref{teff_logg_comp}. 
When water condenses, water vapor is depleted
above the cloud tops causing a decrease at altitude in the gas-phase
abundance of water.  Water clouds form in the atmosphere of an
isolated 1 $M_J$ object within 100 Myr, and within 2 Gyr they form in
the atmosphere of a 12 $M_J$ object. However, at
supersaturations of 1\% and for particle sizes above 10 $\mu$m, such
clouds (and the corresponding water vapor depletions above them) only
marginally affect the calculated emergent spectra. For wavelengths
long-ward of 1 $\mu$m, the cloudy spectra differ from the no-cloud
spectra by at most a few tens of percent. Below effective temperatures
of 160 K, NH$_3$ clouds form. This is likely well below the effective
temperature we can hope to detect.

We expect that planets detected in reflected light will comprise only
a negligible fraction of our sample, and therefore we have not
attempted a thorough treatment of this problem. Rather, we assume a
Bond albedo of 0.4. This assumption would gives rise to errors if the
observing band were coincident with a strong CH$_4$ band or when the
target it is a so-called Class III or ``clear'' extrasolar giant
planet, so named because they are expected to be too hot (\teff $\geq$
350 K) to contain any principal condensates \citep{Sudarsky2000}.

\subsection{Core accretion}

The standard theory for the formation of gas giant planets is the
core accretion model (e.g., \citealt{Pollack1996}), which begins with
dust particles colliding and agglomerating within a protoplanetary disk to
form icy and rocky planetary cores. If the core becomes massive enough
while gas remains in the disk, it can grow by gravitational
accretion of this gas. Gas giants accrete most of the gas within their
tidal reach, filling the Hill sphere around them with a hot, extended,
gaseous envelope. Further accretion is slowed by the dwindling supply
of local raw materials and by the extended envelope, leading to growth
times of 5--10 Myr. The predicted planet formation time is
uncomfortably long compared to the observed $\sim$ 3 Myr lifetime of
protoplanetary disks. Two factors may alleviate this time-scale
problem: 1) inward migration can bring giant planets to fresh,
gas-rich regions of the disk; 2) dust sedimentation may reduce
atmospheric opacity, which leads to more rapid escape of accretion
luminosity and shrinkage of the envelope. \citet{Hubickyj2005} have
shown that reducing the grain opacity to a level observed in L dwarfs
\citep{Marley2002} reduces the planet growth time scale to $\sim$ 1
Myr, well within the lifetime of protoplanetary disks. Once accretion
stops, the planet enters the isolation stage and the planet contracts
and cools at constant mass.

The configuration of the protoplanet at the end of run-away gas
accretion represents the initial conditions for subsequent cooling and
contraction. \citet{Marley2007} and \citet{Fortney2008} have
conducted preliminary calculations that describe the cooling and
contraction of a young planet as it emerges from its parent disc. The
implication of the \citet{Marley2007} results is that giant planets formed by
core accretion are less luminous
post-accretion than had been previously expected because significant
energy is radiated during the formation process. The fully
formed planet has a smaller radius at young ages than hot start models
predict \citep{Burrows1997, Baraffe2003}, leading to a lower
post-formation luminosity. There are two significant observational
consequences: 1) there is a period of very high luminosity ($\sim$
10$^{-2}$ L$_\odot$) which lasts $\sim$ 40,000 yr; 2) the initial
conditions for subsequent evolution are not ``forgotten'' for a time
of order the Kelvin-Helmholtz timescale, which lasts tens of
millions of years. These factors imply that observations of class II
(0.5--3 Myr) and III ($\sim$ 10--100 Myr) young stellar objects afford the
opportunity to probe the planet formation events. The run-away
accretion spike is likely to be broader and fainter than in these
idealized calculations because of gradual accretion across the gap
that the protoplanet forms, and the probability of witnessing this
event in a typical T Tauri star may be much larger than a few percent

The approximations used to compute the luminosity history of a planet
formed by core accretion are similar to those adopted in early studies
of protostellar formation---the runaway gas accretion phase is 
described using the formalism of \citet{Stahler1980}, where matter 
falling onto the central object passes through a 1-d, optically thick shock. 
The 1-d geometry forces all the matter to pass through this
shock, whereas in nature the accretion is at least 2-d and more
realistically 3-d. Comparison with star formation suggests that some
accretion may occur on viscous time scales in a disk rather than on
the fast dynamical timescale depicted in \citet{Fortney2008}. Ultimately, the
virial theorem must be satisfied, and half the gravitational potential
energy is radiated: the \citet{Marley2007} calculations and the hot start
models \citep{Burrows1997, Baraffe2003} represent limiting
cases in the contraction history. 

Despite the preliminary nature of the \citet{Marley2007} results, one
message is clear: the luminosity of young exoplanets encodes
key information about how they were formed. Simulations 
suggest that the timescales for relaxation are longer
for more massive planets, and are at least approximately related
to the Kelvin-Helmholtz timescale, with
\begin{equation}
\tau \sim \frac{G M^2}{R L},
\end{equation}
where $G$ is the gravitational constant, $M$ is the mass of the planet,
$R$ is the radius of the planet, and $L$ is the luminosity of the planet.
This timescale ranges from tens to hundreds of millions of years 
for gas giants, meaning observations of stars with ages in this range will 
yield information about planet formation. Young targets are also the most
promising targets because a brighter planet is more easily detected, 
but planets formed via core accretion
may be cooler at young ages than hot start planets. If planets are
significantly cooler at young ages than hot start models assumed, 
the detection rate of young gas giants will be lower than expected. 

\section{Observation planning}
\subsection{Planet populations \& Monte Carlo simulations} \label{mcpops_sect} 
We use the results of the sensitivity calculations 
described in Section \ref{noise_sect} and the planet formation and atmosphere 
models described in Section \ref{model_sect} to perform Monte Carlo
simulations characterizing the results of different surveys
using the Gemini Planet Imager \citep{Graham2002, Graham2007b, Graham2007a}.
The sensitivity of the instrument to a planet depends 
upon the angular separation between the host star and 
planet and the brightness of the host star. 
The spectrum of a planet depends on its effective temperature and surface
gravity, which themselves are functions of the mass and age of the planet. 
Specifying the mass, age, and orbital elements of a planet and the
brightness and distance of its host star is then sufficient
to determine whether that planet is detectable according to the model used.
By generating distributions of the orbital elements and mass of exoplanets, 
and assuming that planet and host star formation are coeval, we may
then simulate the likelihood of detecting a planet around some star.

\begin{figure*}[!htb]
\begin{center}
\begin{tabular}{p{8 cm} p{8 cm}}
\includegraphics[scale=0.4]{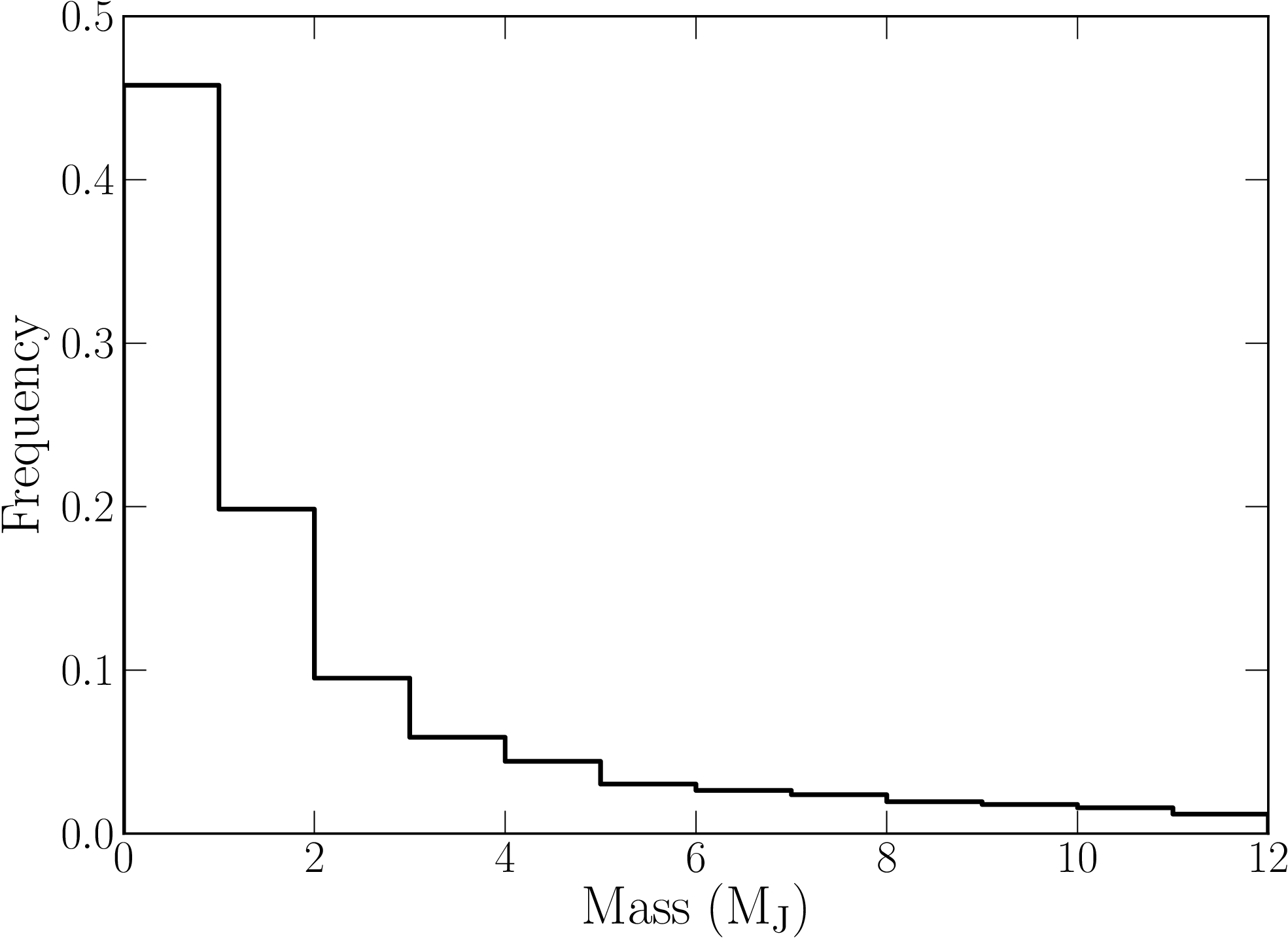} & 
\includegraphics[scale=0.4]{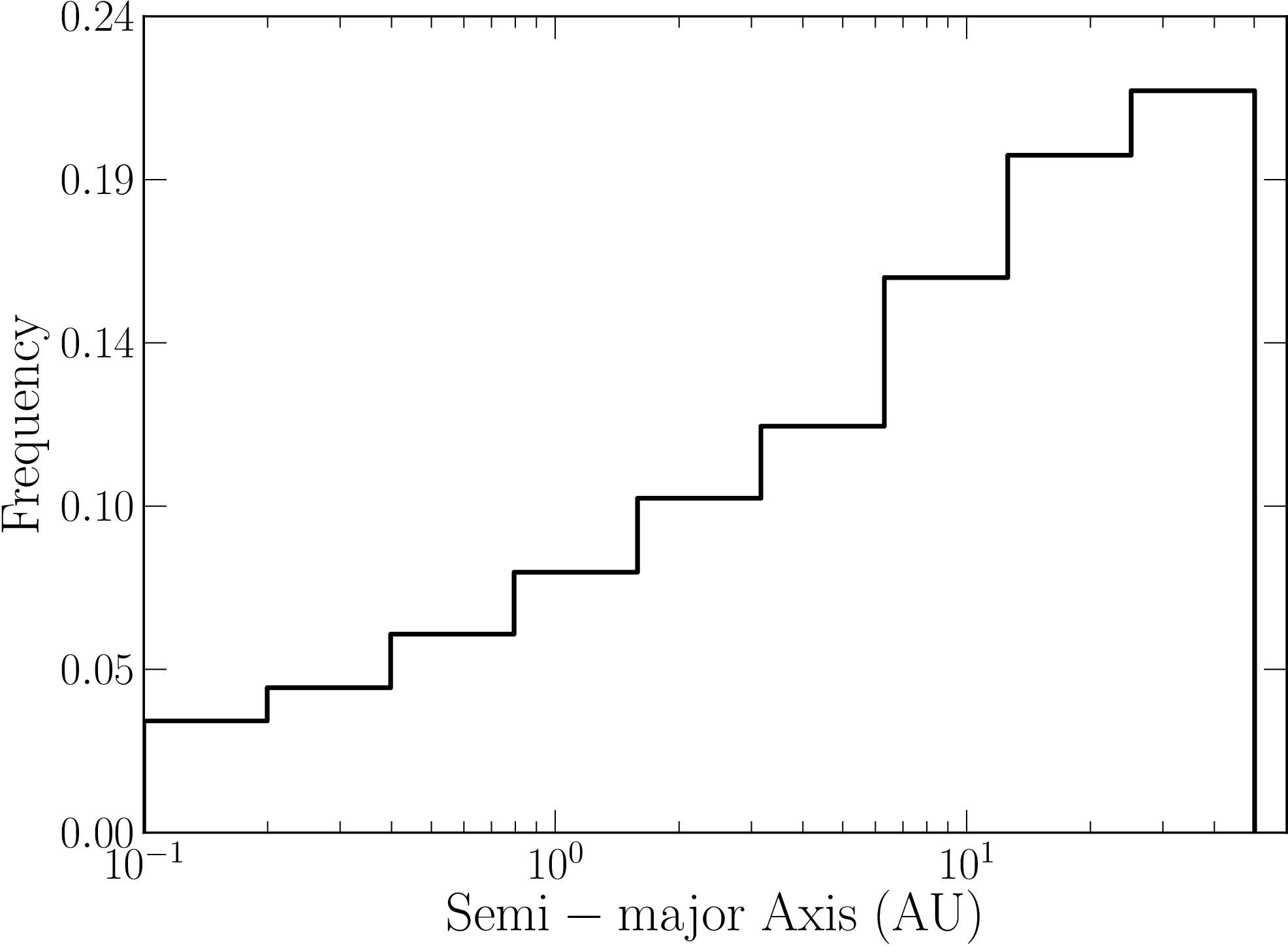} \\
\includegraphics[scale=0.4]{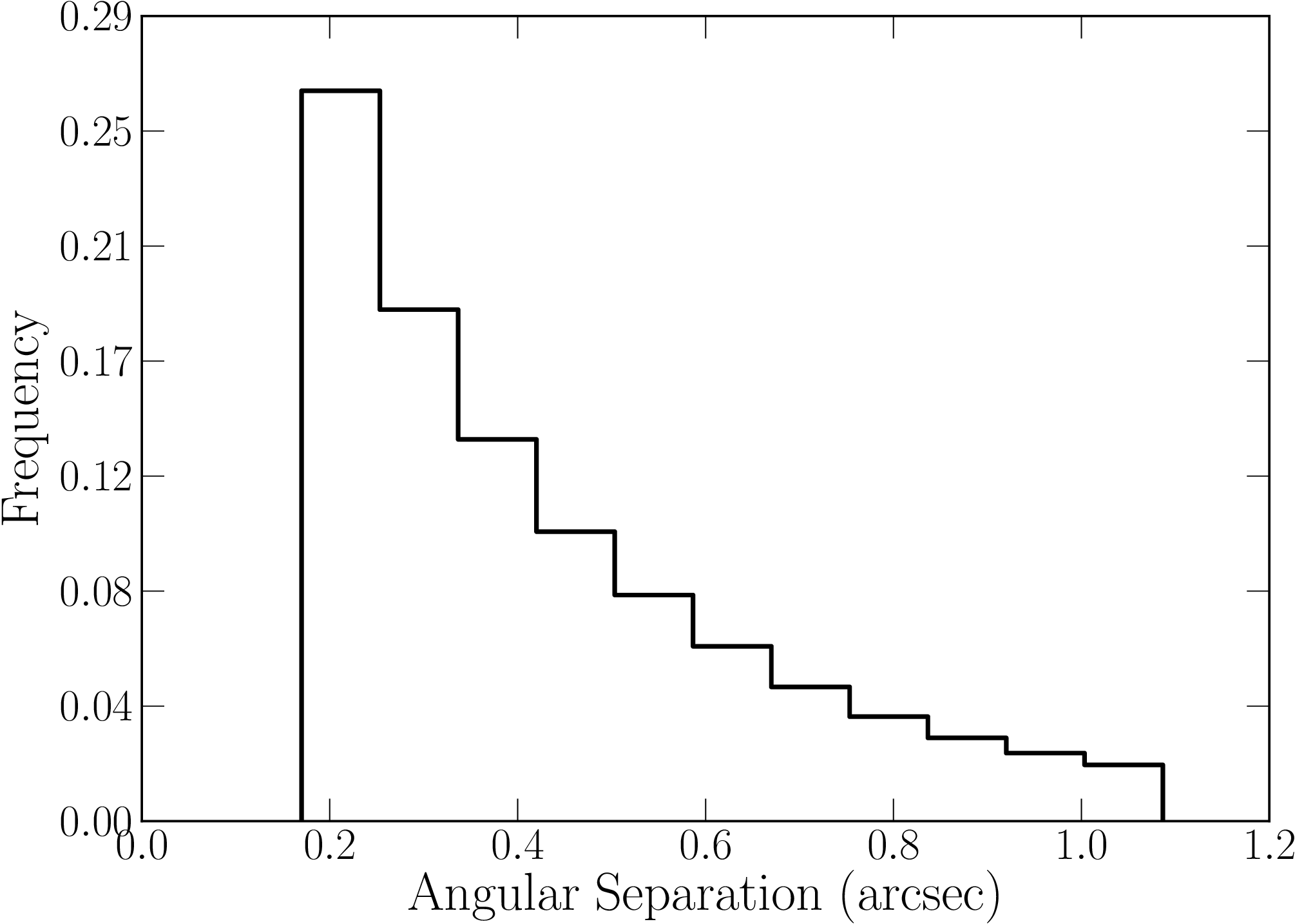} & 
\includegraphics[scale=0.4]{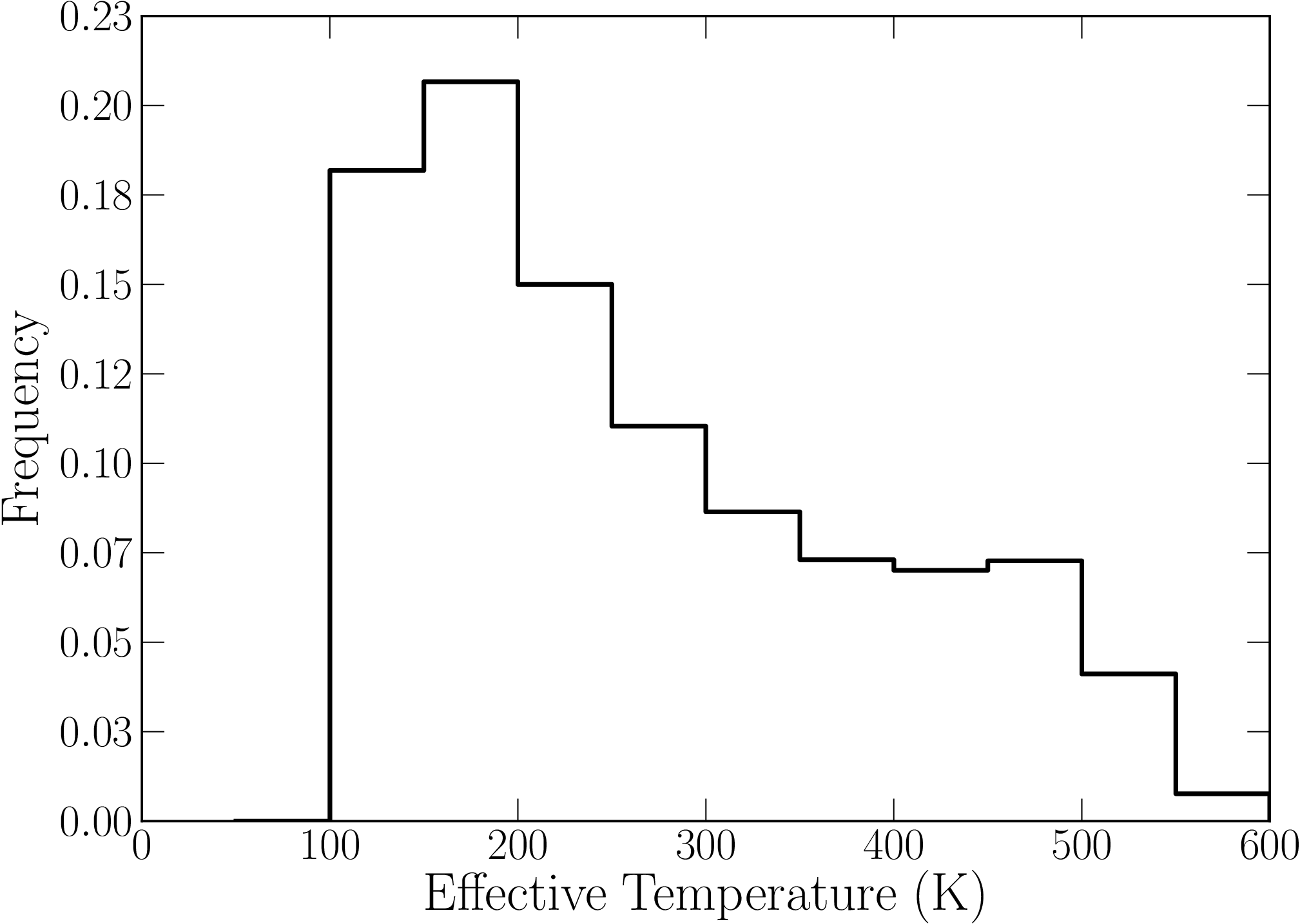} \\
\end{tabular}
\caption{Distributions of planet properties in a typical survey. The mass
and semi-major axis distributions are drawn from extrapolations of 
distributions consistent with radial velocity detected planets. The
angular separation distribution assumes a population of host stars
distributed randomly in a volume within 100 parsecs of the Sun. The
effective temperature distribution assumes a distribution of stellar
ages based on a simple galactic disk star formation rate, and that
planet mass and age map to effective temperature as described in 
\citet{Marley2007}. } \label{input_hists}
\end{center}
\end{figure*}

Lacking an accepted paradigmatic theoretical model for the distributions 
of these parameters, we rely on the distributions at semi-major axes smaller
than $\sim 5$ AU found via radial velocity surveys, and extrapolate
these distributions to larger semi-major axes. The applicability of the
mass and eccentricity distributions found for close-in companions at
large semi-major axes is unclear, as is the extension of the
semi-major axis distribution to larger semi-major axes. Nevertheless,
as a baseline we apply recent results that indicate that the planet
mass and semi-major axis each follow a power law distribution,
with $dN/dM \propto M^{-1.31}$ and $dN/da 
\propto a^{-0.61}$ \citep{Cumming2008}. For the mass, we adopt a
lower cutoff of 0.5 $M_J$, below which a planet is so dim as to make
direct imaging unlikely, and below which radial 
velocity surveys are incomplete.
We take as an upper cutoff 12 $M_J$, reflecting the transition from
planet to brown dwarf at the deuterium fusion limit, thereby 
ignoring any details of how the planet formed.
A histogram of planet masses sampled from this distribution is shown
in the upper left panel of Figure \ref{input_hists}.

Similarly, we adopt cutoffs for the the semi-major axis distribution
of 0.1 AU and 75 AU. Radial velocity surveys show a rising power law for
the semi-major axis distribution in log space out to $\sim 5$ AU, 
and \citet{Nielsen2010} placed an upper limit of 75 AU  
on the semi-major axis distribution quoted in \citet{Cumming2008}.
A lower cutoff may be justified, but already detected exoplanet
systems, such as Fomalhaut, HR 8799, and 1RXS J160929 indicate that
some planets form at large semi-major axes. Such a cutoff is also
consistent with the typical sizes of T Tauri disks \citep{Isella2009}.
A sampled population from this distribution is shown in the upper
right panel of Figure \ref{input_hists}.

The eccentricity distribution is also drawn from radial velocity
observations. Over a decade ago, an eccentricity distribution of
$dN/de \propto e^{-0.5}$ was proposed by \citet{Heacox1999}, but to our
knowledge, no updated eccentricity distribution has been published. We
found that a simpler linear fit to eccentricities adequately
describes observations to date, and adopted that, as results are 
insensitive to the eccentricity distribution anyway.
The rest of the orbital elements are easily
obtained. The mean anomaly and the argument of perihelion are both
drawn from a uniform distribution from 0 to 2$\pi$. The
inclination of the orbit relative to the plane of the sky is drawn
from a distribution of randomly oriented orbits, i.e. $dN/di =
\frac{1}{2}\sin(i)$.

With these distributions, we use a pseudo random number generator and 
the rejection method to generate a population of exoplanets, which 
may be placed around some star. The distribution of masses is then 
converted to distributions of surface gravity, radius, and temperature 
according to the adopted formation model, and the temperature and 
surface gravity then give the emergent spectrum of the planet,  
as described in Section \ref{model_sect}. From this, an expected flux
at the telescope in each spectral channel is calculated. Per the discussion
in Section \ref{matched_sect}, these channels may be summed with our without
weights to get a total flux across the band, but in all results here
channels are summed without weights. Finally, the 
distributions of orbital parameters yield a distribution in angular 
separation. For a given star, we then have relative brightnesses as a 
function of angular separation, which can be compared with the expected 
performance of GPI. The probability of detecting a planet around a star, 
assuming a single planet with mass $0.5 M_J < M < 12 M_J$, is then 
directly obtained from the percentage of relative brightnesses across
the entire wavelength band that are above 
the threshold of detectability, which is taken to be a signal to noise
ratio of 5.

Known exoplanets suggest that these distributions should 
depend on the properties of their host star. The observed metallicity
dependence for short period planets has already been discussed 
in Section \ref{intro_sect}, but will not be included in any
simulations due to uncertainty in its applicability at large semi-major axes.
Doppler surveys show that the 
fraction of stars with planets increases with stellar 
mass over the range from M to A stars \citep{Johnson2008}. 
Microlensing results for the distribution of planets beyond the ice line
are in terms of the mass ratio distribution $dN/dq$, where $q = M_p/M_s$,
rather than a simple distribution of masses $dN/dM$. The success 
at finding planetary mass companions around A stars with other direct 
imaging efforts also supports the importance of host star mass, even 
at large semi-major axes (e.g., Fomalhaut and HR 8799). \citet{Gorti2009} 
predict that the disk lifetime is relatively constant for $M \leq 2 M_\odot$, 
which is also consistent with the notion that Jupiter mass planets are
more common around A stars than they are around M stars. 
To explore the impact of host star mass on direct imaging surveys, we 
tested both a mass ratio distribution $dN/dq$ and a simple mass
distribution $dN/dM$. The mass power law index given in \citet{Cumming2008}
was also used for the mass ratio distribution, 
so that $dN/dq \propto q^{-1.31}$.
The upper and lower cutoffs for $q$ are set such that the lower mass cutoff 
is 0.5 $M_J$ and the highest mass is 12 $M_J$, for the same reasons those
cutoffs are adopted for the $dN/dM$ distribution. 
This is unsubstantiated quantitatively, but seems reasonable qualitatively.
For the rest of the paper, we note when results
are for a mass ratio distribution.

Due to incompleteness, Doppler surveys have not yielded an absolute 
planet hosting probability. Nevertheless, 
\citet{Cumming2008} estimate that 17--20\% of FGK stars host planets 
with $M > 0.3 M_J$ and with $a < 20$ AU.
Surveys using Spitzer IRAC and MIPS find debris disks
around ~15\% of solar-type stars younger than 300 Myr \citet{Carpenter2009} 
and around ~33\% of A-type stars \citet{Su2006} younger than 850 Gyr, 
suggesting a lower limit of this order for the probability of hosting a 
planet. We generally sidestep the uncertainty in planet hosting probabality by
reporting the detection rate in terms of the ratio of the number of planets 
detected to the total number of planets. When reporting the total
number of planets detected around a sample of a fixed number of stars, we
adopt the simple assumption that all stars host a single planet more massive
than 0.5 $M_J$ in the range 0.5--70 AU. 
\subsection{Simulated surveys}
\begin{figure*}[!hbt]
\begin{center}
\includegraphics[scale=0.4]{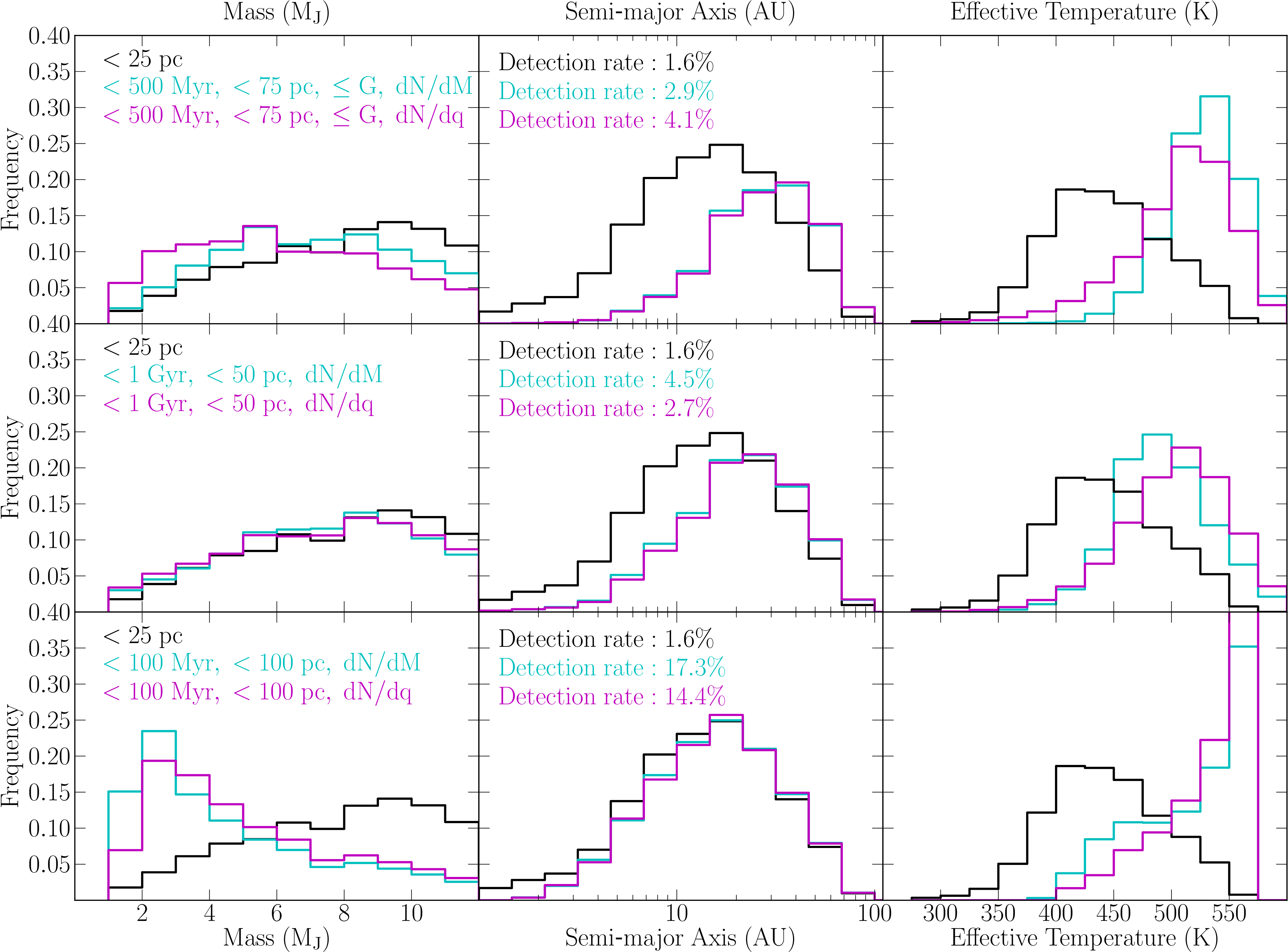}
\caption{Each row of panels plots the intrinsic mass, semi-major axis, and
temperature distributions of detected planets from three different 
surveys. The target selection criteria for the three surveys in each row 
are given by the matching color text
in the left column, and the total detection rate for the survey is given
in the middle column.
The black line is the same in all rows, showing a volume limited selection.
Within each row, the cyan and magenta lines represent the same target selection
criteria, but are distinguished by
whether planet mass depends upon host mass. For cyan lines, planet
mass is independent of host mass, while planet mass scales proprotionally to
host mass for the magenta lines. Cyan and magenta lines
are for a survey emphasizing massive stars in the top row, a survey
emphasizing nearby stars in the middle row, and a survey emphasizing 
young stars in the bottom row.
}
\label{panel_hists}
\end{center}
\end{figure*}

Simulating a survey is a straightforward extension of simulating the 
likelihood of detecting a planet around a single star. Two
of the authors independently wrote code to do this, and used slightly
different approaches. In one approach, a sample 
population of planet properties was drawn from the distributions
described in the previous section and then placed it around each star in some 
sample of stars. In the other approach, a new  
sample population of planet properties was created for each star. The results of
these two approaches were in excellent agreement, providing confirmation
that neither code contained major errors, and the choice of using the
same planet property distribution for each star or different ones 
was unimportant to the results for simulated surveys with a 
large number of planets.

The stellar sample used in a simulation may be either a list of real stars, 
or a Monte Carlo population, where stellar masses are generated from a 
Kroupa IMF, stellar ages are based on a simple galactic disk 
turn-on and turn-off, and stellar distances are generated to match the 
observed stellar density in the solar neighborhood 
\citep{Kroupa2001, Mihalas1981}. 
In either case, the same procedure for simulating the likelihood of detecting
a planet around a single star may be used for a number of stars. 
For different stars, the primary factors that change are the angular separation
on the sky of star and planet, the brightness of the star, and the brightness
of a planet of given mass, as planet brightness depends 
upon the age of the system. 
A record is kept of the planets that are detectable around each star. 

Performing a simulation over a sample of stars provides a total detection
rate and a distribution of properties for detected planets. 
These may be used to compare surveys emphasizing different target selection,
which may then inform where target identification efforts should be focused. 
With this in mind, we simulated a population of stars within 100 parsecs, and
then performed three surveys for samples where targets were selected based on 
different sets of criteria. These criteria were age, distance, and spectral
type (or mass), and the results are shown in Figure \ref{panel_hists}.
To provide a fair comparison between the surveys, 
the criteria adopted for each survey were 
chosen to yield roughly the same number of available targets in each survey. 
The distributions plotted assume complete information
about the planet detected, which will obviously not be the case for real
planets; while the semi-major axis and effective temperature 
may be derived from observations directly, determining the mass of a planet
from direct observations will generally require planet formation and
atmosphere models, or dynamical constraints for multi-planet systems or
planets in disks.  

The results in Figure \ref{panel_hists} show that
observing young stars that are even reasonably near the Sun may be
expected to yield the highest detection rates, as well as the most information
about planets at the lower end of the mass distribution. The primary
drawback in observing the youngest stars is that it requires observing
stars at greater distances. The inner edge of the ``dark hole'' region 
is roughly 0.2\arcsec, so only planets outside of 20 AU may be observed for
a star that is 100 parsecs distant.
If gas giants are very rare at semi-major axes greater than 20 AU or so, 
selecting only young stars would lead to a significantly diminished 
detection rate. Selecting
for stars that are somewhat older and nearer is not as successful in detecting
planets for the semi-major axis distribution that we have assumed, as 
the gas giants with masses $\sim M_J$ and ages $\sim$ Gyr are too dim 
to detect in an hour observation in most cases. Such a survey choice does
have the advantage that the majority of detected planets are at semi-major
axes $<$ 20 AU, and
the existence of gas giants at 5--20 AU is somewhat better established
than the existence of gas giants from 30--70 AU. Present efforts to identify 
targets for GPI are focused primarily on young stars, but  
very nearby stars that also happen to be only moderately young are promising
targets as well. The desirability of early type stars depends on how
the planet mass distribution depends upon stellar mass. Under the assumption
that the planet mass distribution is independent of stellar mass, later type
stars have the highest planet detection rate, since for the same age and
planet mass, the ratio of planet brightness to star brightness will be higher
for dimmer stars. Planet mass very likely does depend on host star mass, and 
making the simple assumption that planet mass scales linearly with host star
mass, planet detection rates are highest for early type stars.  
This result suggests a continued emphasis on identifying early type stars will
be valuable to a successful direct imaging survey.

\begin{figure}[!htb]
\begin{center}
\includegraphics[scale=0.4]{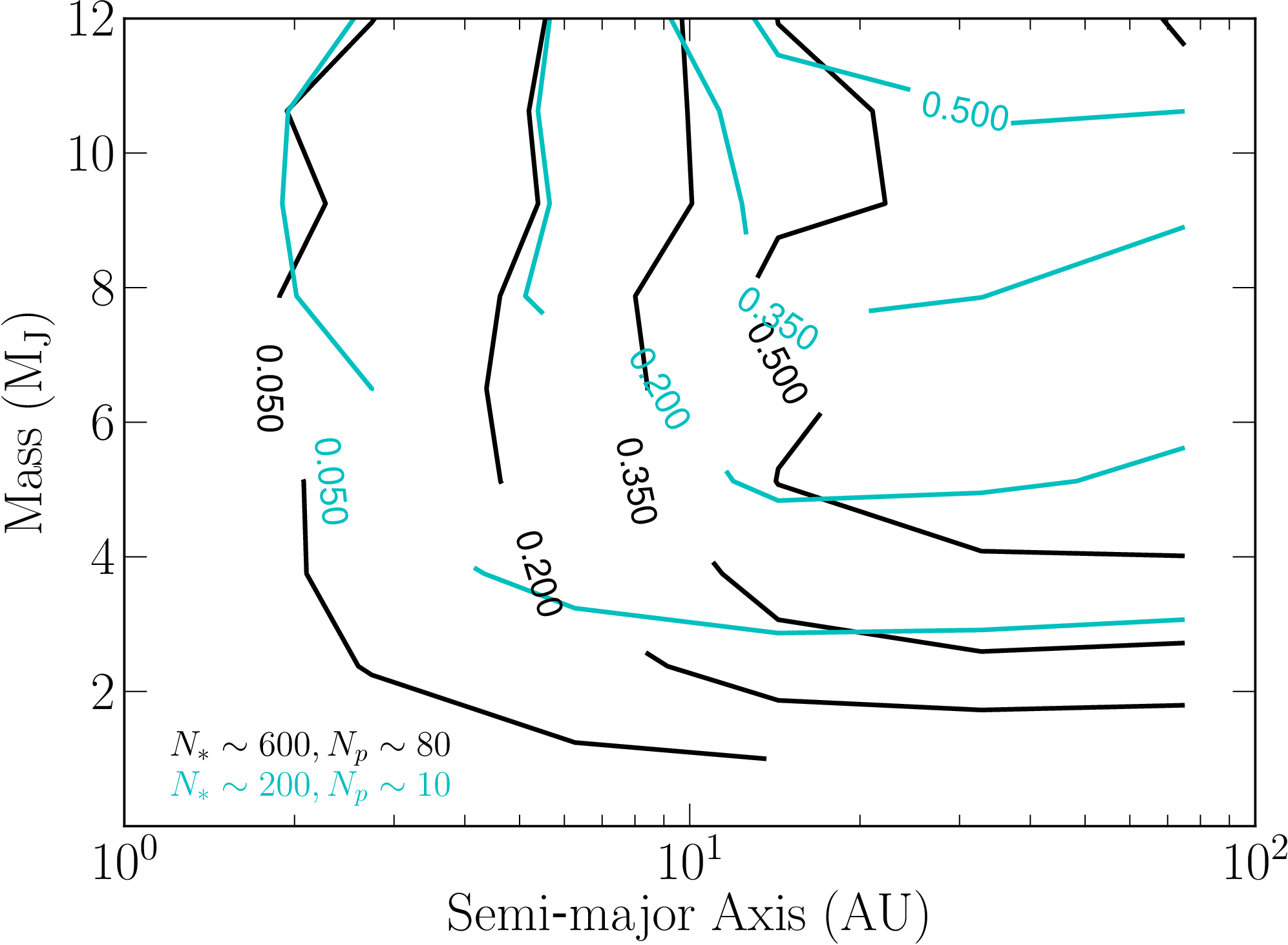}
\caption{Completeness diagram comparing two surveys that assume planets 
evolve according to the models of \citet{Marley2007}.
The cyan represents detection rates for a survey of 
stars within 50 parsecs and younger than 1 Gyr 
target stars from the Geneva-Copenhagen 
Survey \citep{Holmberg2009}, while the black lines are for a survey
of young stars (age $<$ 100 Myr) from a target list being compiled for GPI 
(Song and Bessell, private communication). The colors of the text 
correspond to the same surveys, and show the number of stars surveyed $N_*$ 
and the number of detected planets $N_p$, assuming one planet per star.}
\label{completeness}
\end{center}
\end{figure}

Another means of survey comparision is the completeness diagram, which
encapsulates information about both the rate of detection and the properties
of the planets detected, showing for a grid of planet masses and semi-major axes the
fraction of planets that are detected. Two samples are shown in the completeness 
diagram in Figure \ref{completeness}, with one sample limited to stars younger 
than 100 Myr, and the other limited to stars younger than 1 Gyr 
and nearer than $50$ pc. The young star sample was taken from 
catalogs assembled by I. Song and M. Bessell (private communication). 
Recognizing the importance of a large target set of young, nearby stars 
(defined as age $\leq$ 100 Myr and distance $\leq$ 75 pc), Song and Bessell 
initiated a large-scale program in 2007 to identify and characterize young stars. 
There are approximately 200 known nearby young stars 
($\leq$ 75 pc; $\leq$ 100 Myr) in 
the literature (e.g., \citealt{Zuckerman2004b}; \citealt{Torres2006}). 
To find new candidate young stars, Song and Bessell selected about 3000 
bright ($I \leq$ 9 mag.) Tycho-2 stars from ROSAT catalogs with enhanced 
X-ray emission and potential young star kinematics and obtained optical 
echelle spectra of about 2000 high priority targets with the 2.3-m at Siding Spring.
These spectra were used to extract age indicators including Li 6708 $\AA$, 
Ca II HK, H$\alpha$, and $v \sin(i)$ and estimated ages following 
\citet{Zuckerman2004b}, yielding about 200 additional young stars. 
When combined with an unpublished list of nearby young stars 
(400; Song, Zuckerman, and Bessell since 2000) and currently known members of 
nearby young stellar groups (400), there are approximately 1000 distinct 
solar-type stars that are 100 Myr old or younger within 75 pc. 
There are also more than 1000 “adolescent” stars (100--500Myr.)  
Age uncertainties in the age-dating methods used are age dependent 
(smaller for younger ages) and $\pm$5 Myr for the youngest stars ($\sim$10 Myr old) 
and $\pm$300 Myr for the oldest stars (500 Myr).
The sample used here is limited to roughly 600 stars accessible to the Gemini South
telescope, and has a median age of $~50$ Myr and median distance of $~50$ pc. 
The mean age and distance of the sample are both $~20$\% larger than the median. 
The other sample consists of stars from the Geneva-Copenhagen 
survey \citep{Holmberg2009}, which was chosen because it is a publically
available catalog of stars with estimated ages and distances. From the 
Geneva-Copenhagen catalog, we selected stars accessible to Gemini South that 
are younger than 1 Gyr
and nearer to the Sun than $50$ pc, of which there are roughly 200.
The diagram shows that a survey of moderately young and nearby stars will
probe a reasonable fraction of mass and semi-major axis space, but 
only by surveying a very young sample of stars is there 
a significant probability of imaging gas giants with $M \sim M_J$.
There is, however, a reasonable likelihood of detecting 
the more massive gas giants 
around adolescent stars, and detection of the most massive planets will
be limited primarily by the fraction the orbit during which the
planet is in the ``dark hole'' observing region of GPI.

\subsection{Target Ordering}
In simulated surveys, a planet detection probability is recorded for
each star. These detection probabilities are some function of the 
star's age, distance, and spectral type. We find this function can be
approximated reasonably well with a product of power laws when considering
only stars younger than 2 Gyr. For an older sample, the large number
of stars with effectively zero probability of observing a planet can
skew the resulting fit.
Denoting the star's age by $t$, distance $d$, and mass $M$, the 
detection probability $p$ is
\begin{equation} \label{order1}
p = A \left[\log{\left(\frac{t}{{\rm 1\; Myr}}\right)}\right]^\alpha \left(\frac{d}{{\rm 40\; pc}}\right)^\beta \left(\frac{M}{M_\odot}\right)^\gamma,
\end{equation} 
with the parameters scaled to appropriate values to ease fitting, and 
the logarithm of the age being used as the ages for young stars are 
known with precision at the order of magnitude level.
Performing this regression over multiple trials, and assuming that planet
planets are described by the models of \citet{Marley2007} and that 
mass is independent of stellar mass, we find 
$\alpha \sim -5$, $\beta \sim -1.5$, $\gamma \sim -1.5$. 
While the best fit values are sensitive to the average age of stars in a sample,
they emphasize priority in the same general way: the log of the age
is the most important parameter, and the distance and mass of the star are
of roughly equal importance.

\begin{figure}[!htb]
\begin{center}
\includegraphics[scale=0.4]{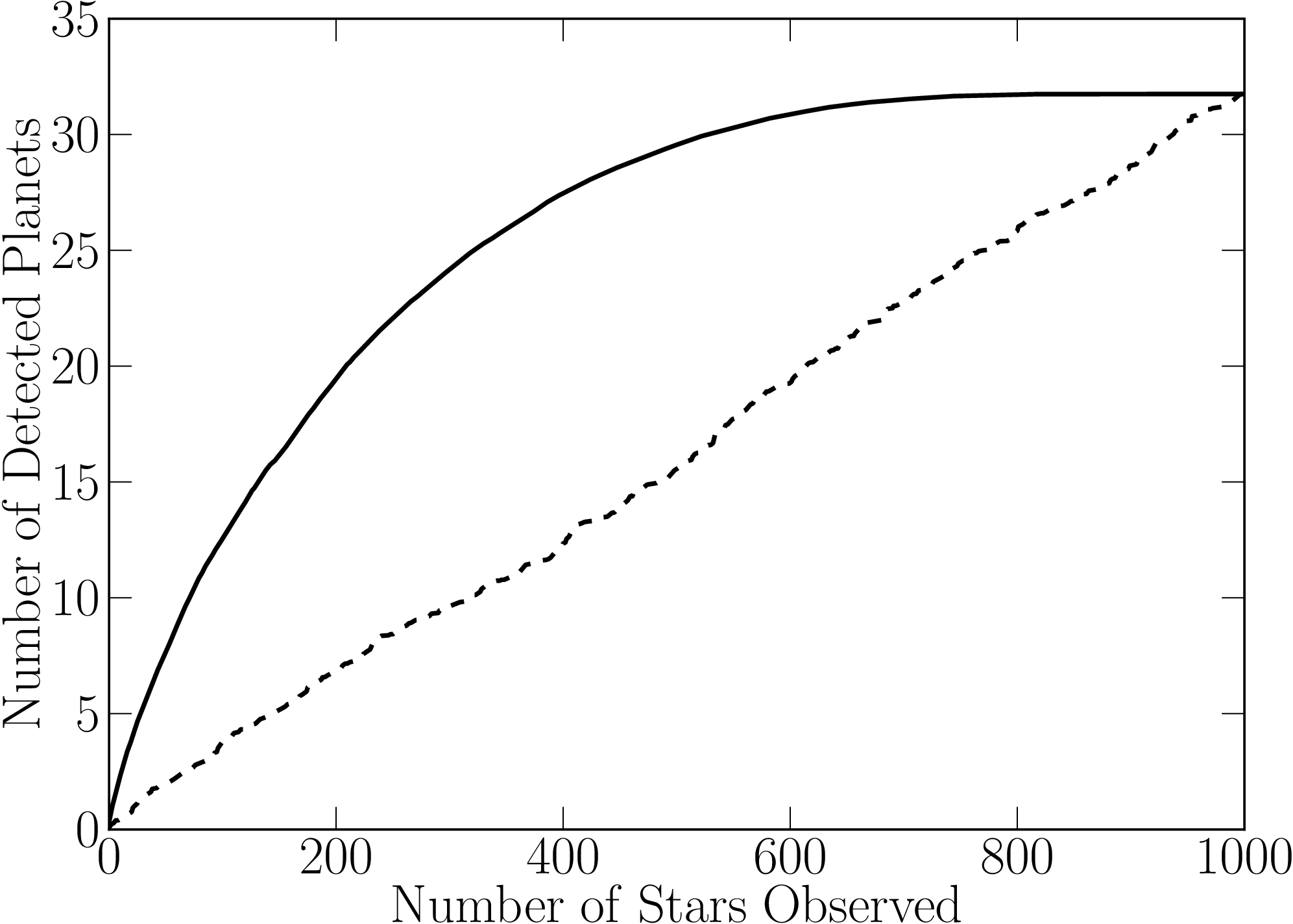}
\caption{The expected number of detected planets is shown for two observing
strategies. The dashed black line indicates targets chosen randomly 
from a sample of stars younger than 1 Gyr and within 70 parsecs.
The solid black line shows the results when the targets are ordered according to
Equation \ref{order1}. } \label{target_ordering}
\end{center}
\end{figure}

Once generated, this detection probability function can be used to 
order target selection before GPI begins observations, as well as for
ordering targets in new simulated surveys. This application is demonstrated
in Figure \ref{target_ordering}, which shows the results of two surveys of 
1000 stars limited in age to be younger than 1 Gyr. If all stars in the
sample are observed, the end result is the same, but when ordering the targets,
approximately 2/3 of detectable planets are found in the first 1/3 of the
sample.

Fitting for the target ordering parameters also gives an at-a-glance
sense of what is important in a survey. As such, it is an interesting
way to test the impact of assumptions about the planet population.
For instance, when the distribution
$dN/dq$ is used rather than $dN/dM$ for generating planet masses, 
detection probability increases with host star mass rather than decreasing.
The other parameters stay roughly the same, with
$\alpha \sim -5, \beta \sim -1.5, \gamma \sim 1$. Likewise,
changing the slope on the semi-major axis distribution unsurprisingly
impacts the importance of stellar distance on a survey.

\section{Expected GPI Performance} \label{perf_sect}
\subsection{Detection rate and distributions}
With 150 nights 
of telescope time, GPI could perform initial 1 hour observations and 
follow-up for roughly 1000 stars. The results of such a survey depend 
strongly on target selection. The predictions of the formation model 
also strongly influence an exoplanet survey, though this effect 
decreases as the mean age of targets increases and model predictions
converge. An age limited survey of stars younger than 1 Gyr
within 80 parsecs will have a detection rate of 4\% for the 
\citet{Marley2007} model, whereas planets that 
evolve according to \citet{Burrows2003} will be detected around 12\% 
of stars. For a survey of the stars younger than 100 Myr being compiled by
Song (2010, private communication) for GPI, and with the important assumption
that the semi-major axis distribution observed for radial velocity exoplanets
continues out to $\sim$ 70 AU, these detection rates increase to 13\% and 
21\% for the \citet{Marley2007} and \citet{Burrows2003} models, respectively.
For a volume limited survey of very nearby stars, the detection 
rate drops to 1--2\% for both models. 

\begin{figure}[!htb]
\begin{center}
\includegraphics[scale=0.4]{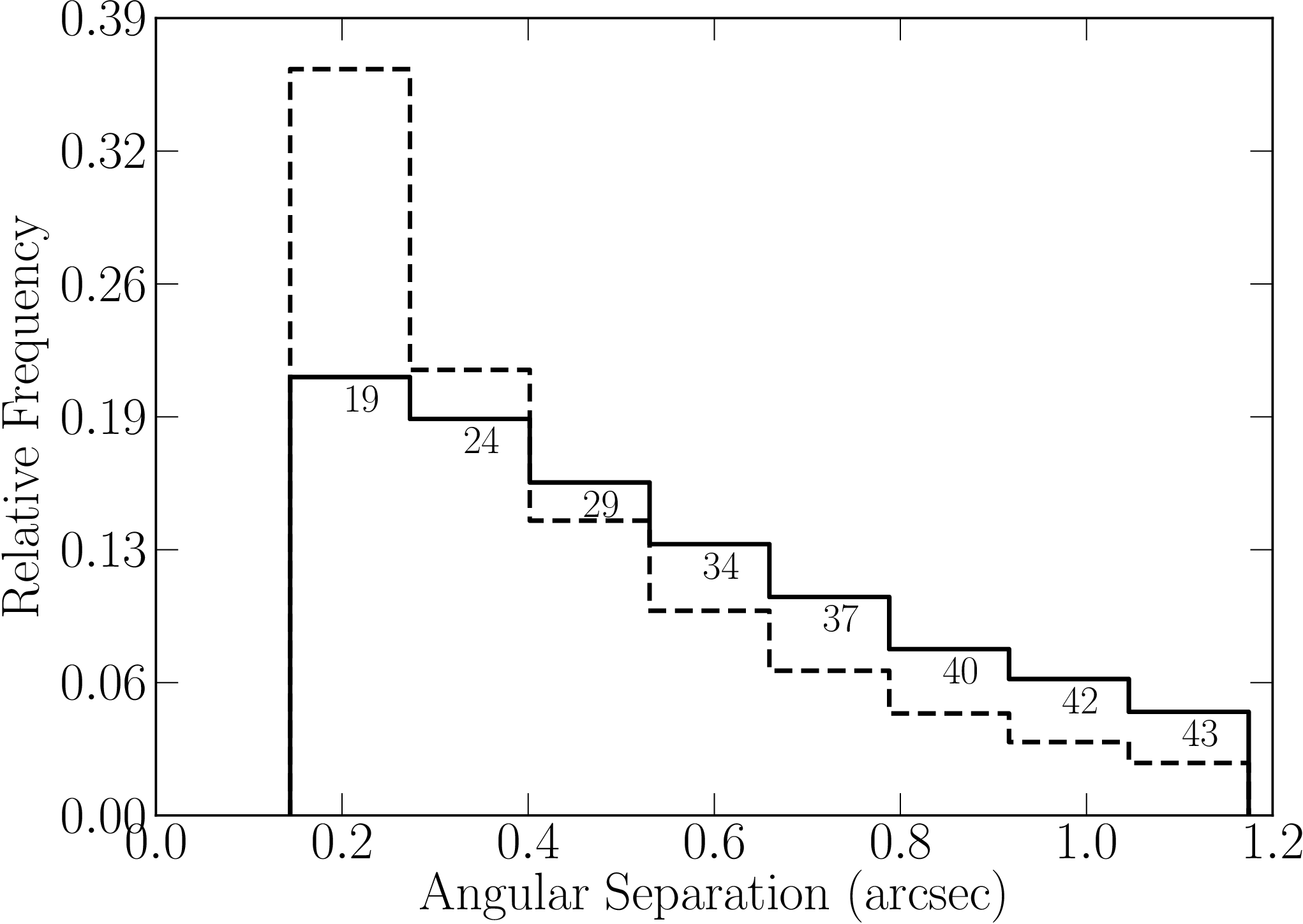}
\caption{The distribution of angular separation for GPI detected planets is 
shown by the solid black line, while the distribution for generated planets is
shown by the dashed black line. The average semi-major axis of planets within 
each bin is printed at the top of the bin.} \label{ang_sep_hists}
\end{center}
\end{figure}

While the existence of a significant population of planets at large
semi-major axes (30--70 AU) will certainly improve 
the detection rate of a GPI survey,
GPI may still be successful so long as gas giants exist at more
moderate semi-major axes.
GPI is capable of detecting planets separated by less than $0.2\arcsec$ from
their hosts. This is illustrated by Figure \ref{ang_sep_hists}, which
shows the distribution of angular separation for planets detected by GPI,
as well as the intrinsic angular separation distribution, given the assumed
semi-major distribution and distribution of stellar distances. The majority of
detected planets will come from the inner half of the GPI ``dark hole.'' 
For the innermost angular separation bin, the 
average semi-major axis of detected planets is 18 AU.
These small separations are where
GPI will be most valuable, probing regions of semi-major axis 
space inaccessible to previous imaging surveys. Figure \ref{ang_sep_hists}
also indicates that regardless of the number
of planets at large semi-major axes, GPI will probe a region of semi-major
axis space which has been unexplored by previous radial velocity surveys
and direct imaging surveys.

Likewise, GPI is capable of detecting a wide range of planet masses. As shown
in Figure \ref{panel_hists}, observations of young stars may be expected
to yield detections of planets with mass $M \sim M_J$ in reasonable numbers. 
For planets with mass $M \sim 8 M_J$ and above and ages less than 1 Gyr,
the detection rate is limited primarily by fraction of an orbit that
planets spend in the ``dark hole'' region. This remains true for objects
beyond the deuterium burning demarcation between planets and brown dwarfs.
Imagining the planetary mass distribution extending to the realm of 
low-mass brown dwarfs up to 30 $M_J$, GPI could expect to find at least 
twice as many brown dwarfs as planets. It should quickly become clear
whether the brown dwarf desert extends to large semi-major axes.

\subsection{Model differentiation}
\begin{figure}[!htb]
\begin{center}
\includegraphics[scale=0.4]{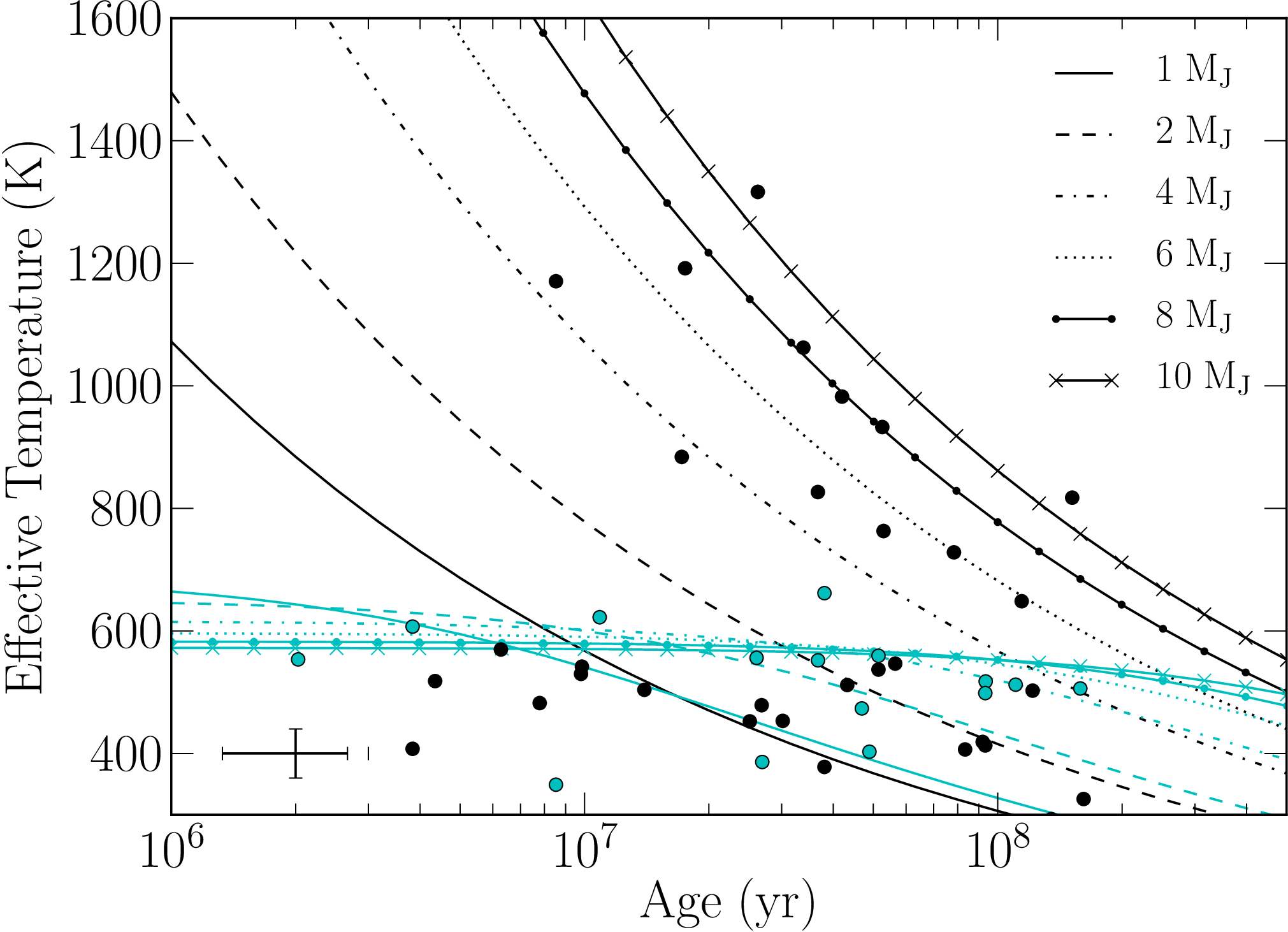}
\caption{Effective temperatures of detected planets are shown as a
function of planet age as circles, with representative error bars in 
the lower left corner. Black
circles indicate planets with properties according to the model of
\citet{Burrows2003}, and cyan circles correspond to planets from the
\citet{Marley2007} model. The lines show expected planet cooling tracks
for each model for planet masses of 1, 2, 4, 6, 8, and 10 $M_J$, 
and the colors for the cooling tracks are black for \citet{Burrows2003} and
cyan for \citet{Marley2007}. More massive planets are at most ages warmer
than less massive planets in each model, but the \citet{Marley2007} planets
all converge to roughly the same temperature at young ages.} 
\label{age_temp_comp}
\end{center}
\end{figure}

The core accretion model of \citet{Marley2007} and the hot start model 
of \citet{Burrows2003} make distinct predictions for the expected 
properties of young gas giants. The differences in 
formation models are greatest for the planets that
are young, so a survey targeting stars with ages $\sim 100$ Myr
and younger is most effective. If gas giants were were
formed via only one of the two scenarios, 
conducting a survey of 100 stars with these ages would be sufficient to
determine which model described the formation of the planets. 
We performed a simulated survey like this twice, once for each model.
Figure \ref{age_temp_comp} shows the expected results of such each survey,
with effective temperatures of detected planets as a function of planet age.
It is clear that the two populations of planets are very different, both to
the eye and from performing simple statistical tests comparing the
distributions and their agreement with each of the models. 
For this simple scenario then, a modest amount of observing time 
can distinguish the two formation models.

In reality, this dichotomy will not exist. The models represent two extreme
cases, and the post formation entropy of any individual planet would likely
fall somewhere in between what each of the models predicts. 
Looking at a distribution of planets will be even more
complicated if planets form via more than one mechanism. It will 
likely be valuable to segregate distributions of detected planets 
by semi-major axis, host mass, and planet metallicity, according to 
predictions about how the planet distribution will depend upon these
stellar properties for each formation mechanism. 
Significant differences in these segregated
distributions could provide useful evidence for a difference in formation
history. 

In simulating formation model differentiation, the other major simplification 
was ignoring the existence of brown dwarfs, which will also likely 
complicate efforts to determine how planets form. Well constrained stellar ages
will be vital in discriminating between planet formation models. 
Objects that are brown dwarfs according to both the deuterium fusion limit and
in the sense of forming like a star are expected to be significantly warmer 
and more luminous than planet mass objects at the same age. Objects with 
masses above the deuterium burning limit that formed via core accretion are
potentially more problematic if their post formation luminosities are
similar to planet mass objects that formed through core accretion. 
However, the surface gravities of objects above and below the deuterium
burning limit will be different enough to be distinguished, as GPI is expected
to achieve a precision of better than 0.2 dex in $\log(g)$.

The HR 8799 system provides an early example of examining planet formation
models using the effective temperatures and ages of detected planets.
As noted in \citet{Marois2008a}, 
the members of the HR 8799 system are not consistent with the simple 
core accretion model presented in \citet{Marley2007}. After the formation
event, planets formed in their model are insufficiently hot 
to match the temperatures of 
the HR 8799 members, even considering the large uncertainty in the age 
of the planets. Given the dynamical constraints on the system that place 
an upper limit on the masses of $\sim 10 M_J$ \citep{Fabrycky2010}, and
despite the fact that the thermal history of a planet formed via 
core accretion is undoubtedly more complicated than the model presented
by \citet{Marley2007}, the temperatures of the HR 8799 planets support
the notion that planets do not form by core accretion at large semi-major
axes.  
The observed abundance of multi-planet systems suggests that
some of the planets GPI will discover will also be in multi-planet systems. 
Finding more systems like HR 8799 will help determine more confidently
how planets form. 

\section{Conclusions}
We have simulated the performance of the Gemini Planet Imager 
in a variety of hypothetical
direct imaging surveys of nearby stars, finding how the detection rate
and properties of detected planets are affected by different survey choices
and assumptions about the exoplanet distribution. These simulations rely upon
calculations and simulations
of the noise properties of the system. They also rely upon models of the 
formation and evolution of gas giant planets and their atmospheres, one
assuming a hot start \citep{Burrows2003} and one assuming a planet
formed by core accretion \citep{Marley2007}. 

Regardless of formation model, detection rates will be highest for
observations of young stars.
For planets that formed via core accretion, roughly 10\% of planets
around stars younger than 100 Myr may be detected, 
and for hot start planets this detection rate may be as high as 25\%.
One major uncertainty is the
frequency of planets at large semi-major axes. If planets are very rare
beyond 30 AU, a GPI survey would be more successful focusing on moderately
young nearby stars, rather than the youngest stars. Only by sampling the
youngest stars though may GPI be expected to 
place significant constraints on the 
low-mass end of the semi-major axis distribution, as the least massive
planets cool the most quickly, and sampling young stars will also be the
most useful approach to better understanding how planets form.

For planets detected with reasonably high signal to noise, GPI will be 
capable of measuring the effective temperature and surface gravity. When
the age of the system is known, these quantities can be 
compared to models of planet formation and evolution. For the idealized
situation where all planets are formed via either the hot start model
of \citet{Burrows2003} or the core accretion models of \citet{Marley2007},
a few tens of detected planets younger than 100 Myr 
would be sufficient to determine the model from which the planets were
drawn. Real planets will not be so simple, but this result suggests that
the characterization of many planets in this fashion will lead to a 
better understanding of how planets form.

We would like to acknowledge Jonathan Fortney, Mark Marley, and Didier Saumon
for helpful discussion regarding their models and for providing additional
model spectra at our request. 

\bibliographystyle{apj}
\bibliography{ms}

\end{document}